\documentclass[usenatbib,usegraphicx]{mn2e}
\usepackage{aas_macros}
\usepackage{rotating}
\usepackage{longtable}
\usepackage{times} 

\def\lsim{~\raise0.3ex\hbox{$<$}\kern-0.75em{\lower0.65ex\hbox{$\sim$}}~}
\def\gsim{~\raise0.3ex\hbox{$>$}\kern-0.75em{\lower0.65ex\hbox{$\sim$}}~}

\def\gs{\mathrel{\raise0.35ex\hbox{$\scriptstyle >$}\kern-0.6em \lower0.40ex\hbox{{$\scriptstyle \sim$}}}}
\def\ls{\mathrel{\raise0.35ex\hbox{$\scriptstyle <$}\kern-0.6em \lower0.40ex\hbox{{$\scriptstyle \sim$}}}}
\newcommand{\arcsecs}{\mbox{$^{\prime\prime}$}}

\newcommand{\arcmins}{\mbox{$^{\prime}$}}
\newcommand{\parcmin}{\mbox{$\stackrel{\prime}{\textstyle .}$}}

\newcommand{\Msolar}{\mbox{$M_{\odot}\,$}}
\newcommand{\Lsolar}{\mbox{$L_{\odot}\,$}}
\newcommand{\degs}{\mbox{$^{o}$}}

\begin{document}

\title[Wide-field mid-infrared and millimetre imaging of 4C\,41.17]
      {Wide-field mid-infrared and millimetre imaging of the
       high-redshift radio galaxy, 4C\,41.17}

\author[Greve et al.]{
\parbox[t]{\textwidth}{
\vspace{-1.0cm}
T.\,R.\ Greve\,$^{\! 1}$, 
D.\ Stern\,$^{\! 2}$, 
R.\,J.\ Ivison\,$^{\! 3,4}$, 
C.\ De Breuck\,$^{\! 5}$, 
A.\ Kov\'acs\,$^{\! 6}$, 
F.\ Bertoldi\,$^{\! 7}$ 
}
\vspace*{6pt}\\
$^1$ California Institute of Technology, Pasadena, CA 91125, USA.\\
$^2$ Jet Propulsion Laboratory, California Institute of Technology, Pasadena, CA 91109, USA.\\
$^3$ UK Astronomy Technology Centre, Royal Observatory, Blackford Hill, Edinburgh EH9 3HJ, UK.\\
$^4$ Institute for Astronomy, University of Edinburgh, Blackford Hill, Edinburgh EH9 3HJ, UK.\\
$^5$ European Southern Observatory, Karl Schwarschild Str. 2, D-85748 Garching bei M\"unchen, Germany.\\
$^6$ Max-Planck Institut f\"ur Radioastronomie, Auf dem H\"ugel 69, 53121 Bonn, Germany.\\
$^7$ Argelander Institute for Astronomy, University of Bonn, Auf dem H\"ugel 71, 53121 Bonn, Germany.
\vspace*{-0.5cm}}

\date{\fbox{\sc Draft dated: \today\ }}
\date{Accepted ... ; Received ... ; in original form ...}

\pagerange{000--000}

\maketitle

\begin{abstract} We present deep 350- and 1200-$\mu$m imaging of the
region around 4C\,41.17 -- one of the most distant ($z$ = 3.792) and
luminous known radio galaxies -- obtained with the Submillimeter High
Angular Resolution Camera (SHARC-II) and the Max Planck Millimeter
Bolometer Array (MAMBO). The radio galaxy is robustly detected at 350-
and 1200-$\mu$m, as are two nearby 850-$\mu$m-selected galaxies; a
third 850-$\mu$m source is detected at 350-$\mu$m and coincides with
a $\sim 2$-$\sigma$ feature in the 1200-$\mu$m map. Further away from
the radio galaxy an additional nine sources are detected at
1200-$\mu$m, bringing the total number of detected (sub)millimeter
selected galaxies (SMGs) in this field to 14. Using radio images from
the Very Large Array (VLA) and {\it Spitzer} mid-infrared (mid-IR)
data, we find statistically robust radio and/or 24-$\mu$m counterparts
to eight of the 14 SMGs in the field around 4C\,41.17. Follow-up
spectroscopy with Keck/LRIS has yielded redshifts for three of
the eight robustly identified SMGs, placing them in the redshift range
$0.5\ls z \ls 2.7$, i.e.\ well below that of 4C\,41.17. We infer
photometric redshifts for a further four sources using their
1.6-$\mu$m (rest-frame) stellar feature as probed by the IRAC bands;
only one of them is likely to be at the same redshift as 4C\,41.17.
Thus at least four, and as many as seven, of the SMGs within the
4C\,41.17 field are physically unrelated to the radio galaxy.  With
the redshift information at hand we are able to constrain the observed
over-densities of SMGs within radial bins stretching to $R=50$ and
$100\arcsecs$ ($\sim 0.4$ and $\sim 0.8\,$Mpc at $z\simeq 3.8$) from
the radio galaxy to $\sim 5\times$ and $\sim 2\times$ that of the
field, dropping off to the background value at $R=150\arcsecs$.  We
thus confirm that 4C\,41.17 resides in an over-dense region of the
Universe, but we have only been able to identify SMGs along the line
of sight to the radio galaxy, typical of the blank-field SMG
population. Finally, we report on the discovery of an extremely
extended ($\sim 110\,$kpc) Lyman-$\alpha$ blob at $z=2.672$ associated
with the brightest 1200-$\mu$m source in the field.
\end{abstract}


\section{Introduction}

Strong evidence has been unearthed by many independent studies,
spanning the entire electromagnetic spectrum, that powerful ($P_{\rm
178MHz}\gs 10^{28}\,h^{-2}$\,W\,Hz$^{-1}$) high-redshift ($z\gs\rm 2$)
radio galaxies (HzRGs) are the progenitors of today's massive
spheroids and giant ellipticals, seen in the early stages of their
formation. Large radio luminosities indicate the presence of a
supermassive black hole, while deep near-IR observations betray large
stellar masses, making HzRGs amongst the most massive known baryonic
systems in the distant Universe (e.g.\ Best, Longair \& R\"{o}ttgering
1998; De Breuck et al.\ 2002; Seymour et al.\ 2007). Another extreme
property of HzRGs is the preponderance of luminous, morphologically
complex Ly$\alpha$ halos, extending up to 200\,kpc from the central
radio galaxies (e.g.\ Reuland et al.\ 2004).  In several cases, strong
H\,{\sc i} absorption has been seen in Ly$\alpha$, suggestive of vast
reservoirs of neutral gas (e.g.\ Hippelein \& Meisenheimer 1993; van
Oijk et al.\ 1997; De Breuck et al.\ 2003; Wilman et al.\ 2004).

The first pointed submillimeter (submm) observations of HzRGs
demonstrated that a large fraction of such systems are extremely
luminous in the rest-frame far-IR waveband ($L_{\mbox{\tiny{FIR}}}\sim
10^{13}\,\Lsolar$) and contain large amounts of dust ($M_d\sim
10^9\,\Msolar$ -- Dunlop et al.\ 1994; Chini \& Kr\"{u}gel 1994;
Ivison et al.\ 1995; Hughes, Dunlop \& Rawlings 1997). Systematic
850-$\mu$m SCUBA surveys of HzRGs by Archibald et al.\ (2001) and
Reuland et al.\ (2004) confirmed these initial findings and showed
that the dust content of HzRGs, as gauged by the 850-$\mu$m
luminosity, increases strongly with redshift:
$L_{\mbox{\tiny{850$\mu$m}}}\propto (1+z)^{3-4}$ out to $z\sim\rm 4$.
This was interpreted as a tendency for HzRGs to host younger stellar
populations at higher redshifts since far-IR luminosities can be
powered by either an AGN or starburst activity but large dust masses
require vigorous recent star formation. Even though HzRGs undoubtedly
host powerful AGNs, starburst activity is therefore known to
contribute substantially to their extreme far-IR luminosities; indeed,
the lack of a correlation between radio power and submm luminosity
seems to exclude a scenario in which the AGN is the dominating power
source (Archibald et al.\ 2001). Furthermore, a handful of HzRGs are
known to harbour large reservoirs of molecular gas ($M(\mbox{H}_2)\sim
10^{11}\,\Msolar$ -- e.g.\ Papadopoulos et al.\ 2000; De Breuck et
al.\ 2003; Greve et al.\ 2004a; De Breuck et al.\ 2005; Klamer et al.\
2005) -- enough to fuel a $\sim$1000\,$\Msolar$\,yr$^{-1}$ starburst
for $\sim$10$^8$\,yr.

The environments of HzRGs are known to be extremely complex and to
harbour a rich variety of galaxies. HzRGs are associated with
over-densities of Ly$\alpha$ emitters (Venemans et al.\ 2007),
Lyman-break galaxies (LBGs -- Miley et al.\ 2004), H$\alpha$ emitters
(Kurk et al.\ 2003), extremely red objects (EROs -- Kurk et al.\ 2004;
Kodama et al.\ 2007), X-ray emitters (Pentericci et al.\ 2002; Smail
et al.\ 2003b), as well as of dusty submm galaxies (Ivison et al.\
2000; Smail et al.\ 2003a; De Breuck et al.\ 2004). One of the most
striking examples of the latter came from the first 850-$\mu$m SCUBA
imaging of HzRGs (Ivison et al.\ 2000; Stevens et al.\ 2003).  This
revealed extended ($\sim$30\,kpc) dust emission around the central
radio galaxies and an apparent over-density of SMGs (by a factor of
two compared to blank-field surveys), suggestive of many vigorous
starbursts occuring within a few Mpc of the central HzRGs.

These observational data all tally with current models of hierarchical
structure formation, in which galaxy formation is anti-hierarchical
and heavily biased towards regions of high density. The various source
over-densities suggest HzRGs are excellent markers of the densest and
most vigorous star-forming regions in the early Universe. HzRGs thus
constitute an important population for benchmarking current models of
galaxy formation, particularly at the high-mass end.  The suitability
of HzRGs for probing galaxy formation and evolution is furthered by
the fact that HzRGs are 1) possible to identify in low-frequency radio
surveys as ultra-steep-spectrum (USS) radio sources (Chambers et al.\
1996; De Breuck et al.\ 2000) which, when combined with near-IR colour
criteria, provide a highly efficient and relatively unbiased (in
particular with respect to their dust properties) selection scheme,
and 2) have accurate radio positions and extreme properties (e.g.\
large rest-frame optical luminosities) which aid their study.

In this paper we present 1200- and 350-$\mu$m maps of the $z=\rm
3.792$ radio galaxy, 4C\,41.17 -- one of the most submm-luminous HzRGs
known (Chini \& Kr\"{u}gel 1994; Dunlop et al.\ 1994; Ivison et al.\
2000; Archibald et al.\ 2001; Stevens et al.\ 2003).  We compare our
data with existing 850-$\mu$m imaging and use deep {\it Spitzer} data
as well as VLA radio imaging to locate mid-IR and radio counterparts
to SMGs in the surrounding field.

Throughout, we adopt a flat cosmology, with $\Omega_m=\rm 0.27$,
$\Omega_\Lambda=\rm 0.73$ and $H_0=\rm
71$\,km\,${\mbox{s}^{-1}}$\,Mpc$^{-1}$ (Spergel et al.\ 2003).

\section{Observations and data reduction}
\label{section:observations-and-data-reduction}

\subsection{(Sub)millimeter data}

A $\sim$58-arcmin$^2$ area, centered on 4C\,41.17 and covering the
entire region previously observed with SCUBA (Ivison et al.\ 2000),
was mapped at 1200-$\mu$m using the 117-channel Max-Planck Millimeter
Bolometer Array (MAMBO -- Kreysa et al.\ 1998) on the IRAM 30-m
Telescope in Granada, Spain.  The 30-m telescope has an effective beam
of 10.7\arcsecs~(FWHM) at 1200-$\mu$m.

The data were obtained during the winter semesters of 2003-2004 and
2004-2005 in excellent weather conditions, when the atmospheric zenith
opacity at 1200-$\mu$m was below 0.25 and the sky noise was low. A
total observing time of 38hr (including calibrations) was obtained for
the project: 9hr in March 2004 and 29hr during the winter 2004-2005.
Standard on-the-fly MAMBO scan maps, each 300$\arcsecs \times$
320\arcsecs~in size, were made at regular grid positions
2\arcmins~apart, thus ensuring uniform coverage across the entire
region. The pointing and focus were checked after each map, i.e.\
every hour or so.

\begin{figure*}
\begin{center}
\includegraphics[width=0.6\hsize,angle=-90]{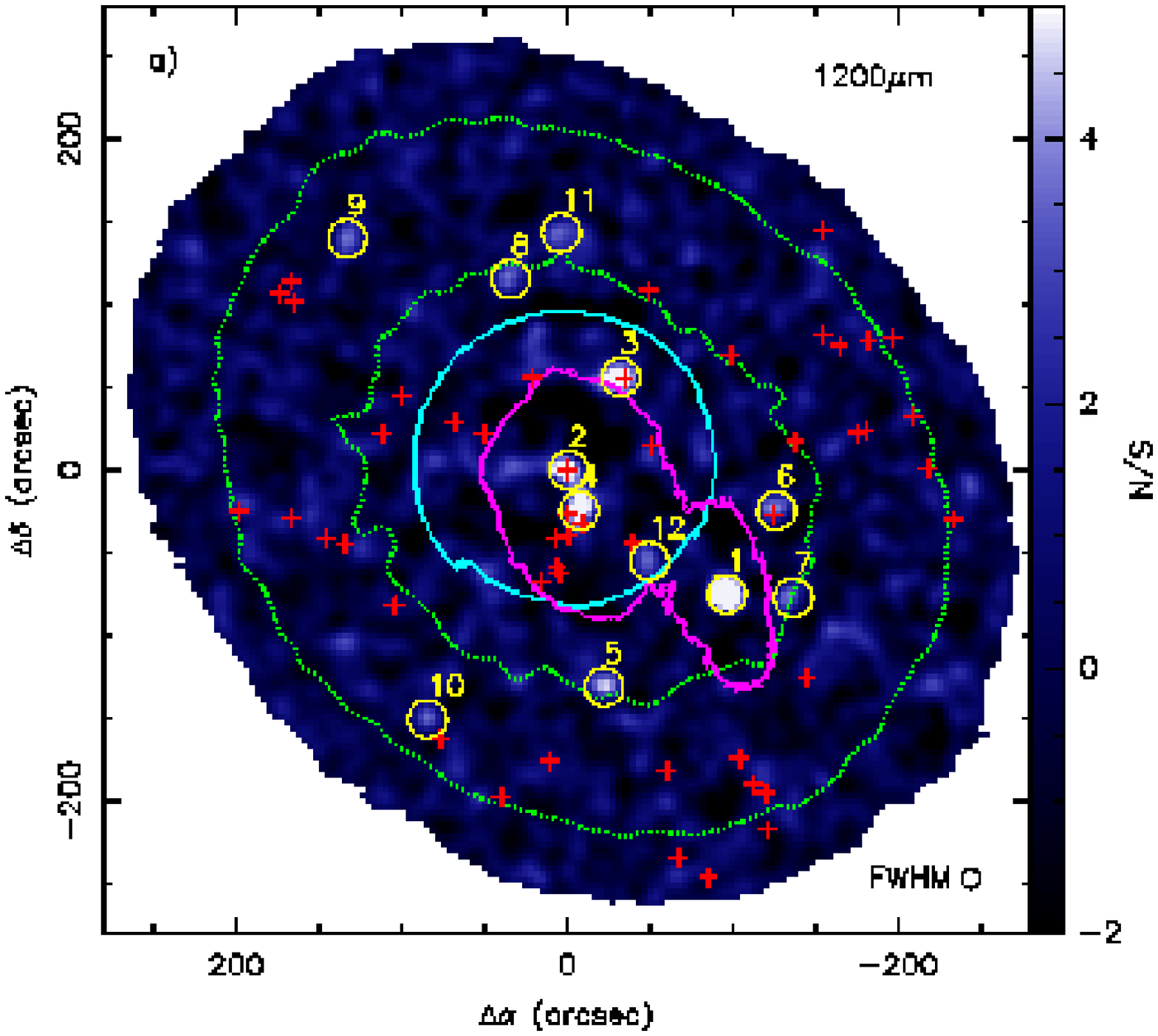}
\includegraphics[width=0.45\hsize,angle=-90]{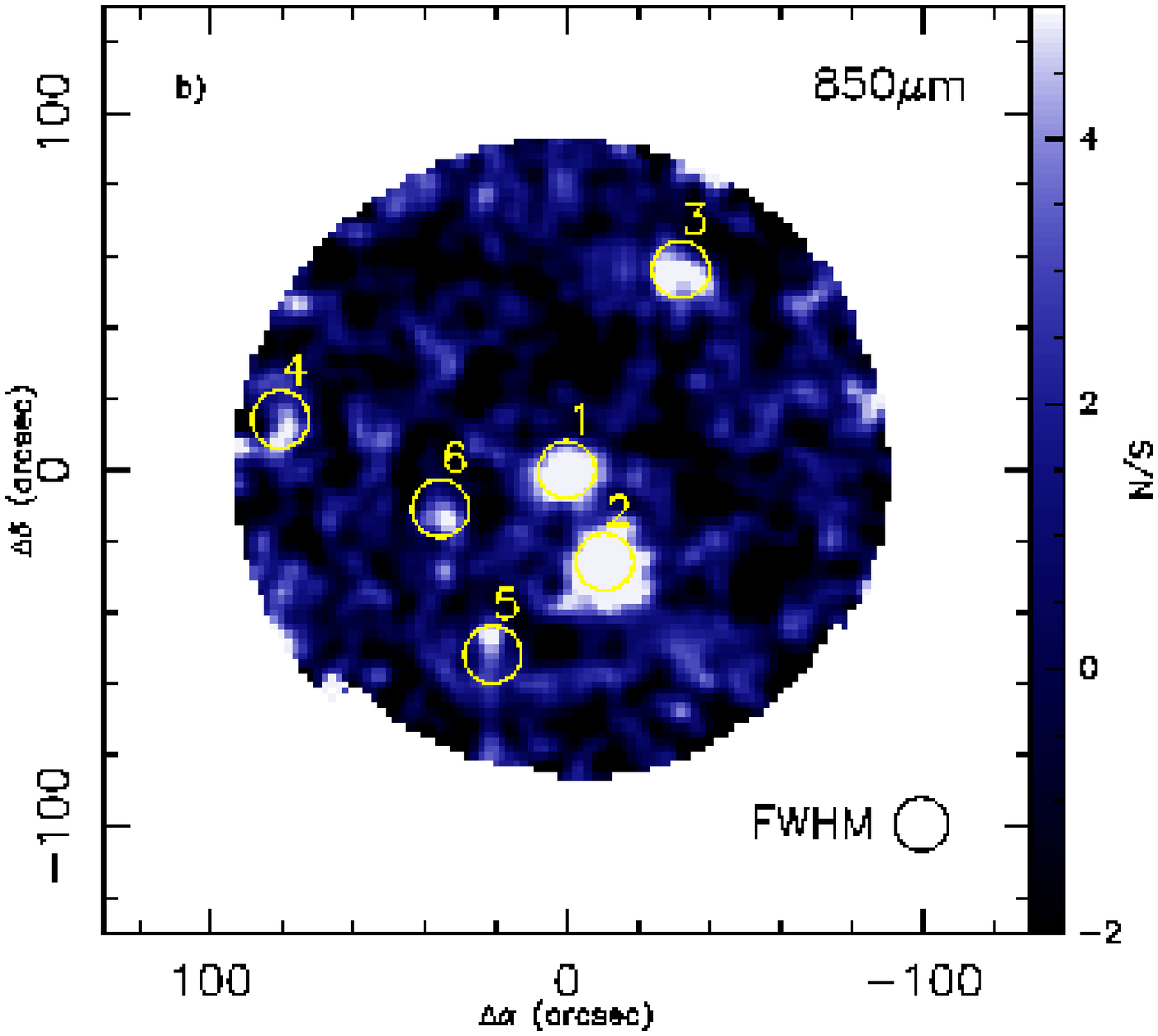}\includegraphics[width=0.45\hsize,angle=-90]{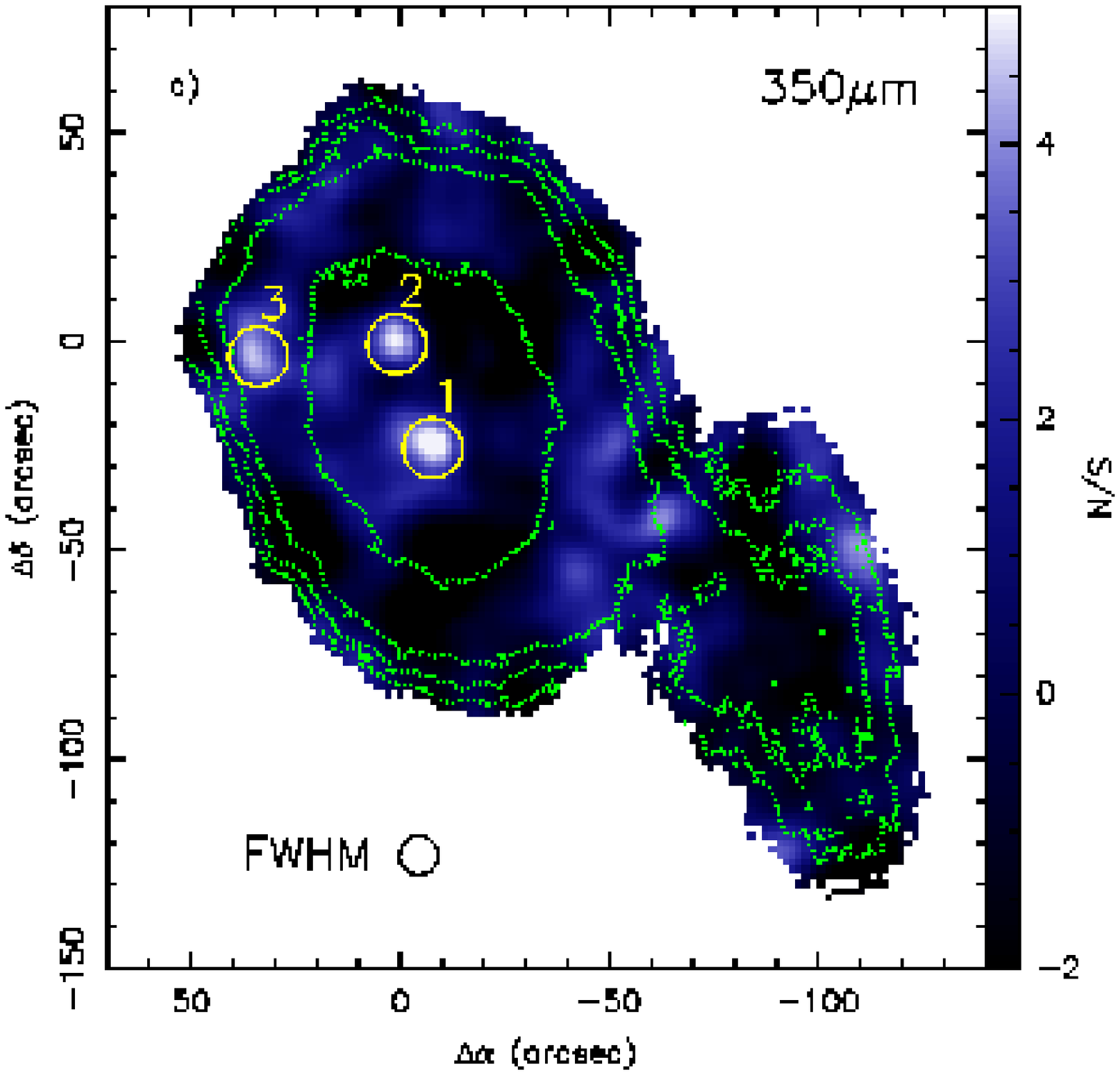}
\caption[]{{\bf a)} 1200-$\mu$m MAMBO signal-to-noise map centred on
4C\,41.17. Green dashed contours mark the 0.5 and
1.0\,mJy\,beam$^{-1}$ r.m.s.\ noise levels across the map. Light blue
and magenta curves indicate the extend of the 850-$\mu$m SCUBA and
350-$\mu$m SHARC-II maps, respectively (Ivison et al. 2000; this
work). Red crosses indicates the X-ray sources detected by Chandra
(Smail et al.\ 2003b). {\bf b)} 850-$\mu$m SCUBA signal-to-noise map
of 4C\,41.17 (Ivison et al.\ 2000). {\bf c)} 350-$\mu$m SHARC-II
signal-to-noise map of 4C\,41.17. The 40 and 60, 70 and
80\,mJy\,beam$^{-1}$ r.m.s.\ noise levels are shown as green dashed
contours. In all three panels we circle sources detected at a
significance $\ge 3.5$-$\sigma$ in yellow; numbering ranks them
according to signal-to-noise ratio (with 1 corresponding the highest
ratio). Also, the axes denote the offset (in arcseconds) from the map
centre R.A.\ 06:50:52.2, DEC.\ $+$41:30:30.9 (J2000). The angular
resolution of the MAMBO, SCUBA and SHARC-II maps are 10.7\arcsecs,
14.7\arcsecs~and 9.2\arcsecs~(FWHM), respectively.  }
\label{figure:mambo-scuba-sharcii-maps}
\end{center}
\end{figure*}

The data were reduced in a standard manner using the {\sc mopsic}
software package (Zylka et al.\ 1998). This involved flagging noisy
bolometers, de-spiking and flat-fielding, as well as removing
correlated sky noise from the time streams of data. Finally, maps were
created on a grid of 4\,arcsec$^2$ pixels.  The final signal-to-noise
map is shown in Fig.\ \ref{figure:mambo-scuba-sharcii-maps}a.

\smallskip

As part of a program to map the most luminous HzRGs at 350-$\mu$m
(Greve et al., in preparation), 4C\,41.17 was observed using the
SHARC-II camera (Dowell et al.\ 2003) at the Caltech Submillimeter
Observatory (CSO) on Mauna Kea in Hawaii. SHARC-II contains a
fully-sampled, pop-up bolometer array: $\rm 32\times 12$ pixels, each
with a size of 4.6\arcsecs, giving a field of view of
2.5\arcmin~$\times$ 0.9\arcmin.  The FWHM beam size is 9.2\arcsec.

Data were taken in excellent weather conditions
($\tau_{\mbox{\tiny{225GHz}}}\ls 0.05$) during a series of observing
runs in 2005 February and May.  Each scan was done in full-power mode
by scanning the array in a 115\arcsec $\times$ 38\arcsec~Lissajous
pattern centred on the central radio position.  The pointing was
checked on an hourly basis using strong submillimeter sources and flux
calibration scans were taken regularly. The absolute calibration error
was found to be less than 30 per cent.  The flux calibration was done
by mapping Uranus once or twice per night and then monitoring a number
of secondary calibrators near the target sources throughout the
night. The pointing model implemented at the CSO during the time of
observations had known systematic trends which we have had to correct
for in the data reduction. This was done by carefully tracking the
residuals between the model and observed pointing sources as a
function of elevation and zenith angle. Using these residuals the
pointing was appropriately updated for each science scan. After doing
so, the typical r.m.s.\ pointing error was of the order
1.5--2.5\arcsec.  The data were reduced using the software {\sc crush}
(Comprehensive Reduction Utility for SHARC-II -- Kov\'{a}cs 2006),
with the {\sc -deep} option ensuring an appropriately aggressive noise
filtering. This is the recommended option for data reduction of faint
and relatively compact sources.  The final map is $\sim
5.2$-arcmin$^2$ in size, covering the central region observed by SCUBA
(Ivison et al.\ 2000) with an r.m.s.\ noise of
$\sim$12\,mJy\,beam$^{-1}$. The resulting 350-$\mu$m signal-to-noise
map of 4C\,41.17 is shown in Fig.\
\ref{figure:mambo-scuba-sharcii-maps}c.

%
%

%
%
\begin{figure*}
\begin{center}
\includegraphics[width=1.00\hsize,angle=0]{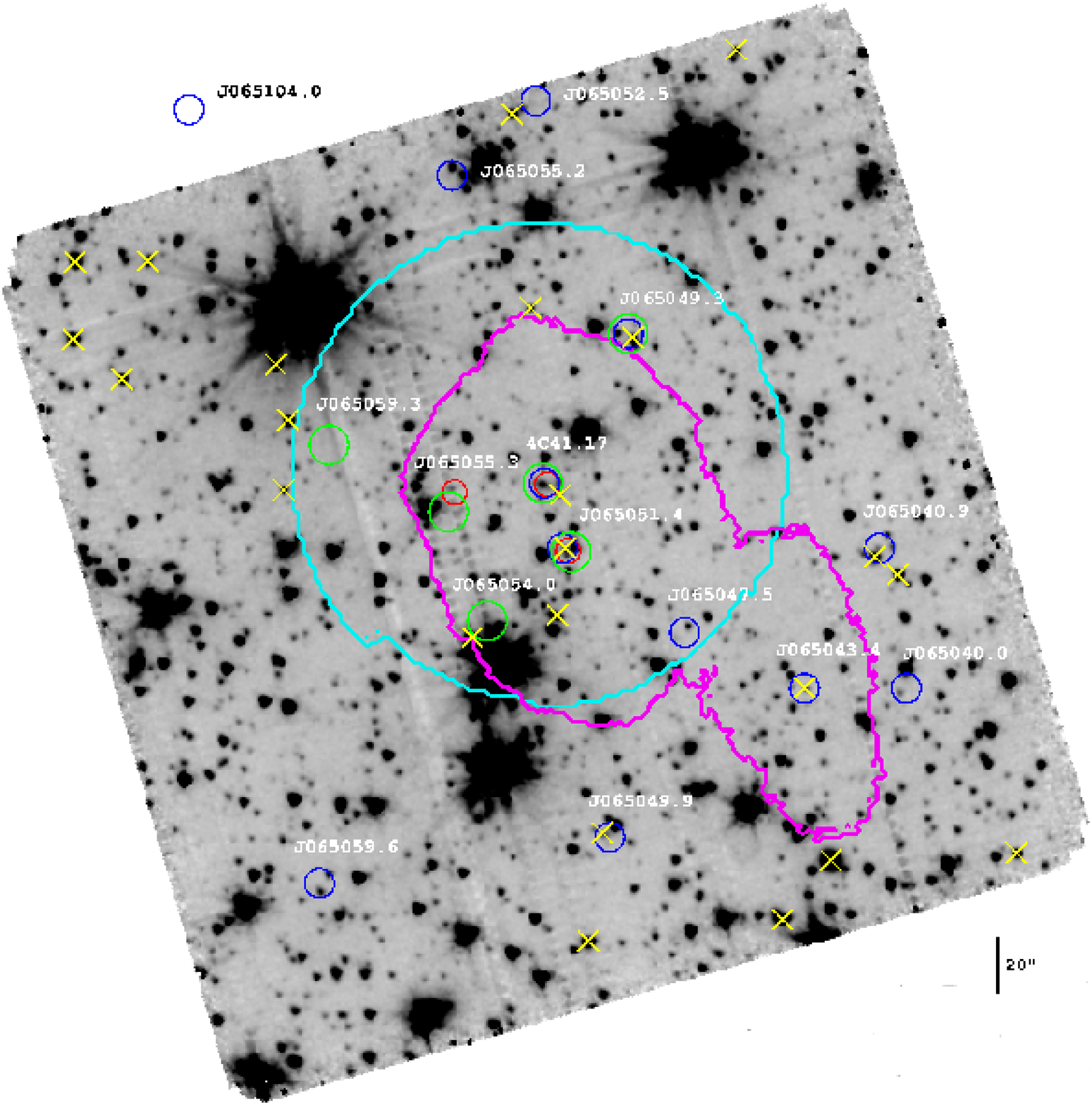}\\
\caption[]{$5\parcmin21\times 5\parcmin21$ IRAC 3.6-$\mu$m image
(grey-scale) of the 4C\,41.17 field. 350-, 850- and 1200-$\mu$m source
catalogues are overlaid as red, green, and blue circles, respectively,
and the diameter of the circles correspond to the FWHM of the CSO
(9.2\arcsecs), JCMT (14.7\arcsecs), and IRAM 30m (10.7\arcsecs) at
those three wavelengths. Sources targeted spectroscopically with
Keck/LRIS are shown as yellow crosses. For clarity, the extent of the
SCUBA and SHARC-II maps have been outlined (light blue and magenta,
respectively). North is up and east is left.  }
\label{figure:combo-map}
\end{center}
\end{figure*}

\smallskip

Finally, we also include in the analysis of this paper the previously
published 850-$\mu$m SCUBA map of 4C\,41.17 (Ivison et al.\ 2000;
Stevens et al.\ 2003). Details of the observations and data reduction
are given in Ivison et al. (2000). Briefly, the map covers a region
$\sim\rm 2.5\arcmins$ in diameter, centred on 4C\,41.17.  The
resolution is $\sim 14\arcsecs$ (FWHM). The 850-$\mu$m signal-to-noise
map is shown in Fig.\ \ref{figure:mambo-scuba-sharcii-maps}b.

\subsection{Radio observations}
Radio data were obtained using the National Radio Astronomy
Observatory's (NRAO\footnote{NRAO is operated by Associated
Universities Inc., under a cooperative agreement with the National
Science Foundation.}) VLA in its A configuration during a period in
2003 July when a correlator fault had ruled out the intended programme
of spectral-line observations (Ivison et al.\ 2007).  A small target
field was required to minimise the impact of bandwidth smearing (due
to the use of continuum mode) so we observed 4C\,41.17 at 1,400\,MHz
for a total of 22.2\,ks over three days, recording two contiguous
intermediate frequency (IF) pairs, each covering 50\,MHz in both left
and right circular polarisations.  Calibration scans of 0646$+$448 and
0137$+$331 were interspersed. Data reduction, imaging and
self-calibration followed standard recipes and the resulting noise
around the central radio galaxy was 23\,$\mu$Jy\,beam$^{-1}$
($1.3\arcsecs$ beam, FWHM), meaning the VLA had achieved a dynamic
range well in excess of 10,000 with little effort.

Further VLA observations were obtained in C configuration during 2005
August -- 5.1\,ks centred on the brightest MAMBO source at 4,860\,MHz,
with additional time on 0644$+$392 and 0137$+$331 for
calibration. After self-calibration the noise level in the resulting
map was 24\,$\mu$Jy\,beam$^{-1}$, with a $3.7\arcsecs$ beam (FWHM).

\subsection{Mid-IR observations}

As part of {\em Spitzer} Cycle 1 IRAC instrument team guaranteed time
observations, the 4C\,41.17 field was imaged deeply with the IRAC and
MIPS cameras. These data were previously reported in Seymour et al.\
(2007).  The data were reduced using an IDL pipeline (see Seymour et
al.\ (2007) for details of the observations and processing). The
formal 3-$\sigma$ point-source sensitivities for the 3.6-, 4.5-, 5.8-
and 8.0-$\mu$m IRAC data were 0.8, 1.1, 3.2 and 4.3\,$\mu$Jy,
respectively. Fig.\ \ref{figure:combo-map} shows the
$5\parcmin21\times 5\parcmin21$ 3.6-$\mu$m image centred on 4C\,41.17.
Of the MIPS data, the 160-$\mu$m images were too noisy to be of any
use, and only the 70- and 24-$\mu$m data were used. The 3-$\sigma$
point-source sensitivity of the 70- and 24-$\mu$m maps were about 1850
and 30\,$\mu$Jy in the deepest parts of the maps.

\subsection{Near-IR imaging and optical spectroscopy}

For near-IR imaging of 4C\,41.17 we used $z$-band observations
obtained during a spectroscopy run with the DEep Imaging Multi-Object
Spectrograph (DEIMOS -- Faber et al.\ 2003) on Keck on UT 2005
November 5, which was forced into imaging mode due to a fault in the
motor which sets the angle of the grating. The total exposure time is
35\,min, obtained as 7 dithered 5-min exposures.

Optical spectroscopy of the mid-IR and radio counterparts (\S
\ref{subsection:counterparts}) to the SMGs detected in the 4C\,41.17
field were obtained with the Low-Resolution and Imaging Spectrograph
(LRIS -- Oke et al.\ 1995) on the Keck Telescope during a run in 2006
January. Remaining space on the LRIS slit masks was used to target
24-$\mu$m emitters with IRAC counterparts.  All spectroscopically
targeted objects are marked by yellow crosses in Fig.\
\ref{figure:combo-map}.

The 6800\,\AA~(D680) dichroic was use to split the light between the
red and blue arm of LRIS. The 400 line mm$^{-1}$ (B400) grism was used
in the blue arm, giving a typical spectral resolution of $\sim
1.1\,$\AA.  For the red arm, we used the 600 line mm$^{-1}$ (R600)
grating, resulting in a spectral resolution of $\sim 1.3\,$\AA.  A
slit width of $1.2\arcsecs$ was adopted. The total integration time
was 1\,hr, split into 20\,min exposures, obtained during dark
conditions in $\sim 1.1\arcsecs$ seeing.  The data were reduced using
the {\sc
bogus2006}\footnote{http://zwolfkinder.jpl.nasa.gov/\~{}stern/homepage/bogus.html}
IRAF package which was specially designed for reducing LRIS slitmasks.

\section{Results}\label{section:results}

\subsection{(Sub)millimeter sources}\label{subsection:submm-sources}

Employing a noise-weighted convolution technique (see Greve et al.\
2004b for details), a total of 12 sources were extracted at a
significance $\ge 3.5$-$\sigma$ from the 1200-$\mu$m MAMBO map. The
1200-$\mu$m source catalogue is listed in Table
\ref{table:mambo-source-list}, and the extracted sources are indicated
in Fig.\ \ref{figure:mambo-scuba-sharcii-maps} as yellow circles. Peak
fluxes as well as aperture fluxes are listed in Table
\ref{table:mambo-source-list}. Flux densities were determined using
apertures 15\arcsecs~in diameter, consistent with the aperture size
used by Ivison et al.\ (2000) to obtain total 850-$\mu$m flux
densities. The same source extraction technique was applied to the
SHARC-II images, resulting in three sources detected at $\ge
4.0$-$\sigma$ significance. The 350-$\mu$m source catalogue is listed
in Table~\ref{table:sharcii-source-list}. Lowering the detection
threshold to $3.5$-$\sigma$ did not result in additional detections.
We list both peak and 15\arcsecs~aperture fluxes in Table
\ref{table:sharcii-source-list}.
%
%
\begin{table}
\scriptsize
\caption{1200-$\mu$m MAMBO sources extracted from the 4C\,41.17 field 
at a significance $\ge 3.5$-$\sigma$.} 
\vspace{-0.5cm}
\begin{center}
\begin{tabular}{lcccc}
\hline
\hline
ID                              &   RA (J2000)  & Dec (J2000)                &   $I_{1200\mu\rm{m}}^a$   &   $S_{1200\mu\rm{m}}^b$  \\
                                &   (h~m~s)     & (\degs~\arcmins~\arcsecs)  &   (mJy/beam)              &   (mJy)                  \\ \hline
MM\,J065043.4$+$412914$^{1)}$   &   06:50:43.4  & $+$41:29:13.8              & $5.1 \pm 0.4$             &   $7.5\pm 0.6$           \\   
MM\,J065052.1$+$413031$^{2)}$   &   06:50:52.1  & $+$41:30:30.8              & $2.8 \pm 0.4$             &   $4.4\pm 0.4$           \\    
MM\,J065049.3$+$413127$^{3)}$   &   06:50:49.3  & $+$41:31:26.9              & $2.8 \pm 0.4$             &   $4.6\pm 0.4$           \\    
MM\,J065051.5$+$413006$^{4)}$   &   06:50:51.5  & $+$41:30:06.4              & $2.3 \pm 0.3$             &   $3.8\pm 0.4$           \\    
MM\,J065049.9$+$412818$^{5)}$   &   06:50:49.9  & $+$41:28:17.8              & $2.3 \pm 0.5$             &   $3.0\pm 0.5$           \\    
MM\,J065040.9$+$413006$^{6)}$   &   06:50:40.9  & $+$41:30:06.4              & $1.9 \pm 0.5$             &   $3.0\pm 0.6$           \\    
MM\,J065040.0$+$412914$^{7)}$   &   06:50:40.0  & $+$41:29:14.0              & $1.9 \pm 0.5$             &   $3.4\pm 0.6$           \\
MM\,J065055.2$+$413226$^{8)}$   &   06:50:55.2  & $+$41:32:26.3              & $1.7 \pm 0.5$             &   $2.6\pm 0.6$           \\    
MM\,J065104.0$+$413251$^{9)}$   &   06:51:04.0  & $+$41:32:50.8              & $2.6 \pm 0.7$             &   $4.3\pm 0.5$           \\    
MM\,J065059.6$+$412800$^{10)}$  &   06:50:59.6  & $+$41:28:00.4              & $2.7 \pm 0.7$             &   $3.5\pm 0.6$           \\    
MM\,J065052.5$+$413254$^{11)}$  &   06:50:52.5  & $+$41:32:54.0              & $1.9 \pm 0.5$             &   $3.2\pm 0.5$           \\
MM\,J065047.5$+$412935$^{12)}$  &   06:50:47.5  & $+$41:29:35.0              & $1.3 \pm 0.4$             &   $2.7\pm 0.5$           \\
\hline
\label{table:mambo-source-list}
\end{tabular}
\end{center}
\vspace*{-0.5cm}
\hspace*{-0.0cm}{\tiny $^a$ Peak fluxes.}\\
\hspace*{-0.0cm}{\tiny $^b$ Total flux within a $D=15\arcsecs$ aperture.}\\
\hspace*{-0.0cm}{\tiny 1) J065043.4; 2) 4C\,41.17; 3) HzRG850.2/J065049.3; 4) HzRG850.1/J065051.4; 5) J065049.9; 
6) J065040.9; 7) J065040.0; 8) J065055.2; 9) J065104.0; 10) J065059.6; 11) J065052.5; 12) J065047.5}\\
\end{table}

%
%
\begin{table}
\scriptsize
\caption{350-$\mu$m SHARC-II sources extracted from the 4C\,41.17 
field at a significance $\ge 3.5$-$\sigma$.} 
\vspace{-0.5cm}
\begin{center}
\begin{tabular}{lcccc}
\hline
\hline
ID                                 &   RA (J2000)  & Dec (J2000)                             & $I_{350\mu\rm{m}}^a$   &   $S_{350\mu\rm{m}}^b$  \\
                                   &   (h~m~s)     & (\degs~\arcmins~\arcsecs)               & (mJy/beam)             &   (mJy) \\ \hline
SMM\,J65051.3$+$413005$^{1)}$      & 06:50:51.3    & $+$41:30:05.3                           & $32 \pm 6$             &   $42\pm 3$             \\
SMM\,J65052.1$+$413030$^{2)}$      & 06:50:52.1    & $+$41:30:30.3                           & $32 \pm 6$             &   $40\pm 3$             \\
SMM\,J65055.3$+$413027$^{3)}$      & 06:50:55.3    & $+$41:30:27.3                           & $43 \pm 9$             &   $47\pm 4$             \\
\hline
\label{table:sharcii-source-list}
\end{tabular}
\end{center}
\vspace*{-0.5cm}
\hspace*{-0.0cm}{\tiny $^a$ Peak fluxes.}\\
\hspace*{-0.0cm}{\tiny $^b$ Total flux within a $D=15\arcsecs$ aperture.}\\
\hspace*{-0.0cm}{\tiny 1) HzRG850.1/J065051.4; 2) 4C\,41.17; 3) HzRG850.5/J065055.3}\\
\end{table}

With the 1200- and 350-$\mu$m data presented here, and the 850-$\mu$m
data of Ivison et al.\ (2000) at our disposal, we are in a unique
position to confirm or reject many of the 850-$\mu$m sources
apparently associated with 4C\,41.17 as first reported by Ivison et
al.\ (2000).  To this end, we make use of Fig.\
\ref{figure:combo-map}, where we have overlaid the 350-, 850- and
1200-$\mu$m source catalogues on the IRAC 3.6-$\mu$m image. Clearly
the three brightest sources detected by Ivison et al.\ (2000) at
850-$\mu$m (4C\,41.17, HzRG850.1 and HzRG850.2\footnote{From this
point on we shall refer to HzRG850.1 and HzRG850.2 as J065051.4 and
J065049.3, respectively.}) are all detected at 1200-$\mu$m at a
significance $\ge 5$-$\sigma$ (Table
\ref{table:mambo-source-list}). Of a further three submm sources
(HzRG850.3, HzRG850.4 and HzRG850.5\footnote{We refer to HzRG850.3,
HzRG850.4 and HzRG850.5 as J065059.3, J065054.0 and J065055.3,
respectively.}) reported by Ivison et al.\ (2000), one (J065059.3)
coincides with a $2$-$\sigma$ peak in the MAMBO map while the
remaining two elude detection at the $\ge 1$-$\sigma$ level. At face
value, the lack of 1200-$\mu$m detections for two latter sources
argues that they are spurious. However, due to their very low
850-$\mu$m flux densities ($S_{850}\simeq 2.4-2.8\,$mJy -- Ivison et
al.\ 2000), we would not expect to detect them at our r.m.s.\ noise
level ($\sim 0.5-0.7\,$mJy$\,$beam$^{-1}$) for values of the
Raleigh-Jeans slope of $\beta = 1 - 3$. In summary, of the five
850-$\mu$m sources reported by Ivison et al.\ (2000), excluding
4C\,41.17 itself, we have confirmed the reality of the two brightest
sources with our 1200-$\mu$m map but cannot confirm nor rule out the
reality of the remaining three. Our unbiased recovery rate of SCUBA
sources by MAMBO is therefore 40 per cent, comparable to that found in
blank-field surveys (Greve et al.\ 2004b; Ivison et al.\ 2005).

%
%
\begin{table*}
\scriptsize
\caption{Summary of observed (sub)mm fluxes for SMGs in the 4C\,41.17 
field which have been observed at more than one (sub)mm wavelength. 
All fluxes are measured within a 15\arcsecs~aperture. After 4C\,41.17, 
which is listed first, the sources are listed in order of increasing R.A.. 
The positions are the average (sub)mm positions (\S \ref{subsection:submm-sources}).}
\begin{center}
\begin{tabular}{lllcccccccl}\hline
\hline
Source           & R.A.           & Dec.           & $S_{350\mu m}$      & $S_{450\mu m}$                & $S_{800\mu m}$                & $S_{850\mu m}$                & $S_{1200\mu m}$     & $S_{1300\mu m}$       & $S_{3000\mu m}$       & Ref.    \\ 
                 & (J2000)        & (J2000)        & [mJy]               &  [mJy]                        & [mJy]                         & [mJy]                         & [mJy]               &                       &                       &         \\
\hline
4C\,41.17        & 06 50 52.1     & $+$41 30 30.6  & $40 \pm 3$          & $22.5\pm 8.5$                 & $17.4\pm 3.1$                 & $12.3\pm 1.2$                 & $4.4\pm 0.4$        & . . .                 & $0.3$                 & [1,2,3,4,5]     \\
                 & . . .          & . . .          & $37\pm 9$           & $35.3\pm 9.3$                 & . . .                         & $12.10\pm 0.88$               & . . .               & $2.5\pm 0.4$          & . . .                 & [6,7,3,8]     \\
                 & . . .          & . . .          & . . .               & $3\sigma\le 56$               & . . .                         & $11.0\pm 1.4$                 & . . .               & . . .                 & . . .                 & [7,9]     \\
J065043.4        & 06 50 43.4     & $+$41 29 13.8  & $3\sigma\le 70$     & . . .                         & . . .                         & . . .                         & $7.5\pm 0.6$        & . . .                 & . . .                 & [1]     \\
J065047.5        & 06 50 47.5     & $+$41 29 35.0  & $3\sigma\le 30$     & . . .                         & . . .                         & $3\sigma \le 6$               & $2.7\pm 0.5$        & . . .                 & . . .                 & [1]      \\     
J065049.3        & 06 50 49.3     & $+$41 31 27.1  & . . .               & . . .                         & . . .                         & $8.7\pm 1.2$                  & $4.6\pm 0.4$        & . . .                 & . . .                 & [1,2]     \\
                 & . . .          & . . .          & . . .               & . . .                         & . . .                         & $6.2\pm 1.9$                  & . . .               & . . .                 & . . .                 & [2]     \\
J065051.4        & 06 50 51.4     & $+$41 30 05.8  & $42\pm 3$           & $34.1\pm 9.3$                 & . . .                         & $12.2\pm 1.2$                 & $3.8\pm 0.4$        & . . .                 & $3\sigma \le 3.3$             & [1,2,10]     \\
                 & . . .          & . . .          & . . .               & . . .                         & . . .                         & $15.6\pm 1.8$                 & . . .               & . . .                 & . . .                 & [7]     \\
J065054.0        & 06 50 54.0     & $+$41 29 39.0  & $3\sigma \le 54$    & $3\sigma\le 26$               & . . .                         & $2.8\pm 0.8$                  & $3\sigma \le 1.5$   & . . .                 & . . .                 & [1,7]     \\
J065055.3        & 06 50 55.3     & $+$41 30 23.8  & $47\pm 4$           & $3\sigma\le 26$               & . . .                         & $2.4\pm 0.8$                  & $3\sigma \le 1.5 $  & . . .                 & . . .                 & [1,7]     \\
J065059.3        & 06 50 59.3     & $+$41 30 45.0  & . . .               & . . .                         & . . .                         & $6.5\pm 1.6$                  & $3\sigma \le 1.5$   & . . .                 & . . .                 & [1,7]     \\
\hline
\end{tabular}
\end{center}
\label{table:data}
\vspace*{-0.2cm}
\hspace*{-0.0cm}{\tiny [1] This work; [2] Stevens et al.\ (2003); [3] Archibald et al.\ (2001); 
[4] Hughes et al.\ (1997); [5] De Breuck et al.\ (2005); [6] Benford et al.\ (1999); [7] Ivison et al.\ (2000); 
[8] Chini \& Kr\"{u}gel (1994); [9] Dunlop et al.\ (1994);}\\
\hspace*{-0.0cm}{\tiny [10] D. Downes, private communications.}\\
\end{table*}

Turning to the 350-$\mu$m recovery rate, we see from Fig.\
\ref{figure:combo-map} and Table \ref{table:sharcii-source-list}, that
4C\,41.17 and J065051.4 are detected at 350-$\mu$m at a significance
of $\sim 5$-$\sigma$.  Unfortunately, both J065049.3 and J065059.3
fall outside of the SHARC-II map, making it impossible to say anything
about their 350-$\mu$m properties. J065054.0, however, lies within the
SHARC-II map, coinciding with a negative feature. We derive a
$3$-$\sigma$ upper limit of $S_{\mbox{\tiny{350$\mu$m}}}\le 54$\,mJy,
which is above the 350-$\mu$m peak flux density ($\sim$40\,mJy) one
derives by assuming an extreme spectral index of $\beta = 3$ and a
850-$\mu$m flux of 2.8\,mJy.  Thus, the lack of a 350-$\mu$m
counterpart to J065054.0 cannot be taken as a convincing argument
against its reality as we wouldn't expect a detection given the depth
of the 350-$\mu$m map in this region.  Interestingly, however, we find
a significant ($\gs 4$-$\sigma$) 350-$\mu$m source
(SMM\,J65055.1$+$413027) at a position $\sim 5\arcsecs$ north of
J065055.3.  The high significance of this source and the fact that
emission is seen in two independent maps, each made from half of the
SHARC-II data, leaves little doubt about the reality of this source.
Is it the 350-$\mu$m counterpart to J065055.3, or an unrelated source?
The expected 1-$\sigma$ positional uncertainty in any given direction
is given by $\theta\rm \simeq 0.6 FWHM/SNR$ (Ivison et al.\ 2005,
2007), which in the case of SCUBA and SHARC-II yields $\theta \sim
2.9\arcsecs$ and $\sim 1.2\arcsecs$, respectively. Thus, the combined
positional uncertainty is $\sqrt{2.9\arcsecs^2 +
1.2\arcsecs^2}=3.1\arcsecs$. This is in good agreement with extensive
Monte Carlo simulations by Scott et al.\ (2002), showing that typical
positional offsets of low-signal-to-noise sources detected by SCUBA
are $\gs 4\arcsecs$. We thus conclude that the positional offset
between the two sources is not significant, and we shall therefore
assume that they are the one and same SMG, namely J065055.3.  As was
argued above, the lack of a 1200-$\mu$m detection of this source is
not a strong argument against its reality.  Thus, of the three SMGs
reported by Ivison et al.\ (excluding 4C\,41.17) within the region
covered by SHARC-II, at least one, but more likely two (namely
J065051.4 and J065055.3), have been detected at 350-$\mu$m.

Two sources, namely J065047.5 and J065043.4, are detected at
1200-$\mu$m but not at 350- and/or 850-$\mu$m, despite lying within
one or both of the SCUBA/SHARC-II maps. Of those, the most prominent
source is J065043.4, located $\sim 2.1\arcmins$ south-west of the
radio galaxy, within the region covered by SHARC-II. This source is
detected at $\gs 10$-$\sigma$ at 1200-$\mu$m; with a flux density of
$S_{\mbox{\tiny{1200$\mu$m}}}=7.5\pm 0.4$\,mJy, it is the brightest
MAMBO source in the entire field. Unfortunately, J065043.4 is not
detected at 350-$\mu$m and only a 3-$\sigma$ upper limit of
$S_{\mbox{\tiny{350$\mu$m}}}\le 70$\,mJy can be put on its 350-$\mu$m
flux density, largely due to the fact that the source is located in a
part of the SHARC-II map which received very little integration time
(about 0.5\,hr). Given the brightness of this source at 1200-$\mu$m,
however, we would have expected a 350-$\mu$m detection (for $\beta =
1-3$) even at the low sensitivity available in this part of the map.
In \S \ref{subsection:mambo-blob} we discuss the possibility that this source
may be a flat spectrum radio source, where the emission at 1200-$\mu$m
(but not at 350-$\mu$m) is dominated by radio synchrotron emission
stemming from a buried AGN.  For the other source (J065047.5), which
is the faintest 1200-$\mu$m source in our sample, we put 3-$\sigma$
upper limits on its 350- and 850-$\mu$m peak flux densities of
$S_{\mbox{\tiny{350$\mu$m}}}\le 30$\,mJy and
$S_{\mbox{\tiny{850$\mu$m}}}\le 6$\,mJy, respectively.

Finally, exploring the MAMBO map outside of the region covered by
SCUBA and SHARC-II, we find seven sources detected at significance
$\ge 3.5$-$\sigma$ (Fig.\ \ref{figure:mambo-scuba-sharcii-maps}).  We
note that all of these sources are significantly fainter at
1200-$\mu$m than the sources within the SCUBA/SHARC-II maps, with the
exception of J065047.5 which is the faintest source in the sample.

To summarise this section, we note that the total number of SMGs,
selected at 350-, 850- or 1200-$\mu$m, within the region outlined by
the MAMBO 4C\,41.17 map is 14 (excluding 4C\,41.17 itself).  Of these,
seven (three) have been observed (detected) at two or more (sub)mm
wavelengths.  Specifically, we have confirmed three out of the five
SCUBA sources (not including 4C\,41.17) reported by Ivison et al.\
(2000). Of those, one (J065051.4) was detected at both 1200- and
350-$\mu$m; another (J065049.3) lies outside the SHARC-II map but is
robustly detected at 1200-$\mu$m, while the last source is detected at
350-$\mu$m but not at 1200-$\mu$m. A further two 1200-$\mu$m sources
(J065047.5 and J065043.4) were covered by SCUBA and/or SHARC-II but
were not detected at 850- and/or 350-$\mu$m.

In Table \ref{table:data} we list our measured (sub)mm fluxes for the
seven SMGs (and 4C\,41.17) with observations at more than one (sub)mm
wavelengths, along with all previous (sub)mm data published in the
literature to date.  The latter includes independent 450- and
850-$\mu$m SCUBA observations of 4C\,41.17 (Ivison et al.\ 2000;
Archibald et al.\ 2001; Stevens et al.\ 2003), as well as at
1200-$\mu$m (Chini \& Kr\"{u}gel 1994; this work). In general there is
good agreement between independent flux measurements -- in particular,
we note the excellent agreement between the 350-$\mu$m flux densities
for 4C\,41.17 reported by Benford et al.\ (1999) and ourselves. Where
possible, the listed positions are an average of two or more of the
1200-/850-/350-$\mu$m positions.

\subsection{Radio/mid-IR counterparts to SMGs around 4C\,41.17 and spectroscopic redshifts}\label{subsection:counterparts}

Due to the link between synchrotron radio and far-IR emission in
starburst galaxies (Condon et al.\ 1992), and the advent of radio
interferometers capable of large-field, high-resolution imaging, deep
radio observations are currently the most efficient way of identifying
SMGs. The positions of two-thirds of blank-field SMGs have been
pinpointed in this way (Ivison et al.\ 1998, 2000, 2002a; Smail et
al.\ 1999, 2000). Lacking radio observations (or in the absence of a
radio counterpart), deep near- and/or mid-IR imaging can reveal the
presence of either EROs or mid-IR (24-$\mu$m) sources, all of which
are sufficiently rare objects on the sky that they must be either
direct counterparts to the SMGs or (as is sometimes seen) nearby,
associated objects.

In this section we seek to identify radio, near-IR and/or mid-IR
counterparts to the 14 SMGs extracted from our (sub)mm maps of the
4C\,41.17 field, and to this end we have excised $z$-band, IRAC and
MIPS postage stamps centred on the average (sub)mm position for each
SMG (see \S \ref{subsection:submm-sources}) -- each $30\arcsecs$ on a
side and with 1.4- and/or 4.9-GHz radio contours overlaid. The postage
stamp images are shown in Appendix A (Fig.\ \ref{figure:postamps}),
where we also give a detailed description of each source.

In cases where a radio source is present, the likelihood of it being
the correct identification was calculated employing the same scheme
adopted by Ivison et al.\ (2007), in which a radio source is
considered robust if it is peaking at $\ge 4$-$\sigma$, and has an
integrated flux density in excess of 3-$\sigma$. We adopt these
criteria for both the 1.4- and 4.9-GHz fields, and find the surface
density of such robust sources to be 0.65 and 0.043 per sq.\ arcmin,
respectively. We then search for robust radio sources within
8\arcsecs~of each SMG and calculate the formal significance of each
submm/radio association using the method of Downes et al.\ (1986). To
this end we have used the radio number counts at 1.4- and 4.9-GHz from
Bondi et al.\ (2003) and Condon et al.\ (1984). A submm/radio
association is considered significant if the probability, $P$, of it
happening by chance, corrected for the number of ways that such an
association could have happened by chance, is $P\le 0.05$. The
significance of the mm/radio associations are listed in Table
\ref{table:mips-radio-fluxes}, along with their radio flux densities
and positional offsets from the (sub)mm centroids. Our relatively
shallow radio imaging covers only a small fraction of the MAMBO field
due to small 5-GHz primary beam and the pernicious effect of bandwidth
smearing in our 1.4-GHz continuum observations.  Nevertheless, the
radio ID fraction is commensurate with that found in earlier SMG
studies (Ivison et al.\ 2002a), with statistically significant
counterparts found for two SMGs (J065043.4 and J065051.4). We are not
able to determine meaningful upper limits on the radio flux densities
for the remainder of the SMGs, so only the robust detections are
listed in Table \ref{table:mips-radio-fluxes}.

It is seen from the last column in Fig.\ \ref{figure:postamps} in
Appendix A that at least nine of the 14 SMGs (not including 4C\,41.17)
have an 24-$\mu$m source within a 8\arcsecs~search radius. The same
strategy used to calculate the significance of the mm/radio
associations was adopted to infer the robustness of the mm/mid-IR
counterparts. We used the 24-$\mu$m number counts by Papovich et al.\
(2004) to estimate the surface density of sources at various flux
density levels.  The significance of the mm/mid-IR associations are
listed in Table 4, along with their mid-IR fluxes and offsets. Given
the slightly larger 24-$\mu$m beam size ($\rm FWHM \simeq
3.5\arcsecs$, compared to the $\sim 1\arcsecs$ resolution in the
radio), we also list all 24-$\mu$m sources within 15\arcsecs. We find
that seven of the SMGs have statistically robust 24-$\mu$m
counterparts within 8\arcsecs~of the (sub)mm centroid. One of these,
namely J065051.4, was also identified in the radio. In the case of
J065054.0, the 24-$\mu$m counterpart lies just outside the $8\arcsecs$
search radius yet we find it unlikely to be a random association. We
find two instances, namely J065049.3 and J065055.2, where the
24-$\mu$m emission is the result of two or more sources blending
together, thus complicating the identification process. In both cases,
however, the bulk of the 24-$\mu$m emission can be traced to a single
IRAC source, making this the most likely counterpart. This is
consistent with the findings by Pope et al.\ (2006) that in most cases
where multiple 24-$\mu$m sources are found within a SCUBA beam, only
one -- namely the brightest 24-$\mu$m source -- is dominating the
submm emission.

In total, therefore, we have robustly identified -- either via their
radio or 24-$\mu$m emission - eight of the 14 SMGs detected.
Including the 24-$\mu$m data have more than doubled the number of
robust identifications, thereby illustrating the usefulness of this
band for locating SMGs. Even so, due to the shallowness of our radio
and mid-IR data, the combined radio and 24-$\mu$m recovery rate
obtained is significantly less than the typical $\sim$80 per-cent or
so obtained for blank field SMGs (Pope et al.\ 2006; Ivison et al.\
2007) when extremely deep VLA and {\it Spitzer} data are available.

Of the eight robustly identified SMGs, six were targeted
spectroscopically, namely J065040.9, J065043.4, J065049.3, J065049.9,
J065051.4 and J065054.0.  From these six spectra, we could determine
three robust redshifts ($0.507\le z\le 2.672$), excluding them as
physically associated sources to 4C\,41.17. The remaining three
spectra showed no evidence of any continuum nor line
emission. Appendix A gives further details on the spectroscopic
results.
%
%
\begin{table*}
\scriptsize
\caption{24-/70-$\mu$m MIPS and 1.4-/4.9-GHz radio counterparts to
4C\,41.17 and the 14 SMGs found in its vicinity. The MIPS 24-$\mu$m
fluxes were extracted from a 9.26\arcsecs~aperture using aperture
corrections of 3.052 from the MIPS Data Handbook. The
24-$\mu$m-(sub)mm and radio-(sub)mm association probabilities are
given in boldface in cases where the association is deemed significant
($P\le 0.05$).}
\begin{center}
\begin{tabular}{lcccccccc}
\hline
\hline
ID                 &  $S_{24\mu\rm{m}}$    & 24$\mu$m-(sub)mm     & $P$          & $S_{70\mu\rm{m}}$  & $S_{1.4\rm{GHz}}$   &  $S_{4.9\rm{GHz}}$ & radio-(sub)mm     & $P$ \\
                   &                       & separation           &              &                    &                     &                    & separation        &     \\ 
                   &  ($\mu$Jy)            & (")                  &              & ($\mu$Jy)          &  ($\mu$Jy)          & ($\mu$Jy)          & "                 &     \\ \hline
4C\,41.17          & $366\pm 21$           & 0.8                  & {\bf 0.002}  & $3\sigma\le 2870$  & $2.35\times 10^5$   & $2.53\times 10^4$  & 1.3               & $\mathbf{4\times 10^{-6}}$    \\    
J065040.0          & $133\pm 31$           & 6.8                  & 0.101        & . . .              & . . .               & . . .              & . . .             & . . .    \\    
J065040.9          & $349\pm 31$           & 4.0                  & {\bf 0.029}  & . . .              & . . .               & . . .              & . . .             & . . .    \\    
MIPS-1             & $384\pm 31$           & (11.6)               & (0.088)      & . . .              & . . .               & . . .              & . . .             & . . .    \\    
J065043.4          & $3\sigma \le 73$      & . . .                & . . .        & . . .              & . . .               & $109.2\pm 31.1$    & 0.3               & $\mathbf{9\times 10^{-5}}$         \\    
                   & $300\pm 24$           & 7.4                  & 0.062        & . . .              & . . .               & . . .              & . . .             & . . .    \\    
                   & $163\pm 24$           & (8.7)                & (0.095)      & . . .              & . . .               & . . .              & . . .             & . . .    \\    
J065047.5          & $174\pm 21$           & (11.4)               & (0.057)      & $3\sigma\le 5130$  & . . .               & . . .              & . . .             & . . .    \\    
J065049.3          & $276\pm 27$           & 1.1                  & {\bf 0.006}  & $3\sigma\le 4115$  & . . .               & . . .              & . . .             & . . .    \\    
                   & $214\pm 27$           & 4.7                  & 0.067        & . . .              & . . .               & . . .              & . . .             & . . .    \\    
                   & $214\pm 27$           & (10.5)               & (0.095)      & . . .              & . . .               & . . .              & . . .             & . . .    \\    
                   & $191\pm 27$           & (15.0)               & ($>$0.1)     & . . .              & . . .               & . . .              & . . .             & . . .    \\    
J065049.9          & $155\pm 31$           & 1.1                  & {\bf 0.012}  & $3\sigma\le 2115$  & . . .               & . . .              & . . .             & . . .    \\    
J065051.4          & $448\pm 24$           & 1.0                  & {\bf 0.002}  & $3\sigma\le 1850$  & $114.5\pm 31.4$     & . . .              & 1.0               & {\bf 0.002}    \\    
J065052.5          & $179\pm 31$           & (10.2)               & (0.084)      & $3\sigma\le 3640$  & . . .               & . . .              & . . .             & . . .    \\    
J065054.0          & $604\pm 21$           & (9.0)                & {\bf (0.043)}& $3\sigma\le 5360$  & . . .               & . . .              & . . .             & . . .    \\    
J065055.2          & $245\pm 27$           & 2.9                  & {\bf 0.031}  & $3\sigma\le 4680$  & . . .               & . . .              & . . .             & . . .    \\    
                   & $153\pm 27$           & 3.4                  & 0.060        & . . .              & . . .               & . . .              & . . .             & . . .    \\    
J065055.3          & $393\pm 24$           & (14.4)               & (0.098)      & $3\sigma\le 6155$  & . . .               & . . .              & . . .             & . . .    \\    
J065059.3          & $3\sigma \le 65$      & . . .                & . . .        & $3\sigma\le 6079$  & . . .               & . . .              & . . .             & . . .    \\    
J065059.6          & $167\pm 34$           & 3.0                  & {\bf 0.048}  & . . .              & . . .               & . . .              & . . .             & . . .    \\    
                   & $58\pm 34$            & 5.1                  & 0.089        & . . .              & . . .               & . . .              & . . .             & . . .    \\    
                   & $266\pm 34$           & (14.8)               & (0.079)      & . . .              & . . .               & . . .              & . . .             & . . .    \\    
J065104.0          & $63\pm 34$            & (8.6)                & ($>$0.1)     & . . .              & . . .               & . . .              & . . .             & . . .    \\    
                   & $123\pm 34$           & (8.9)                & (0.078)      & . . .              & . . .               & . . .              & . . .             & . . .    \\    
\hline
\label{table:mips-radio-fluxes}
\end{tabular}
\end{center}
\end{table*}

\section{Analysis}\label{section:analysis}

\subsection{Mid- and near-IR properties}\label{section:NIR-properties}
With 4C\,41.17 being at a redshift of $z=\rm 3.792$ the {\it Spitzer}
observations are sampling the near-IR/optical light coming directly
from the stellar populations and/or the AGN in this system.  If the
near-IR emission is dominated by stellar light we expect to see a
characteristic stellar 'bump' at $\sim$1.6\,$\mu$m (rest-frame) in the
combined Spectral Energy Distribution (SED) from a composite stellar
population, due to the minimum in the H$^-$ opacity at 1.6\,$\mu$m in
the photospheres of cool stars (John 1988). In the case of a strong
AGN contribution, however, the non-thermal emission would give rise to
a power-law SED. Thus, the mere shape of the SED in this portion of
the spectrum can be used as a way of discriminating between
AGN-dominated systems and a significant stellar population.

\begin{figure*}
\begin{center}
\includegraphics[width=1.0\hsize,angle=0]{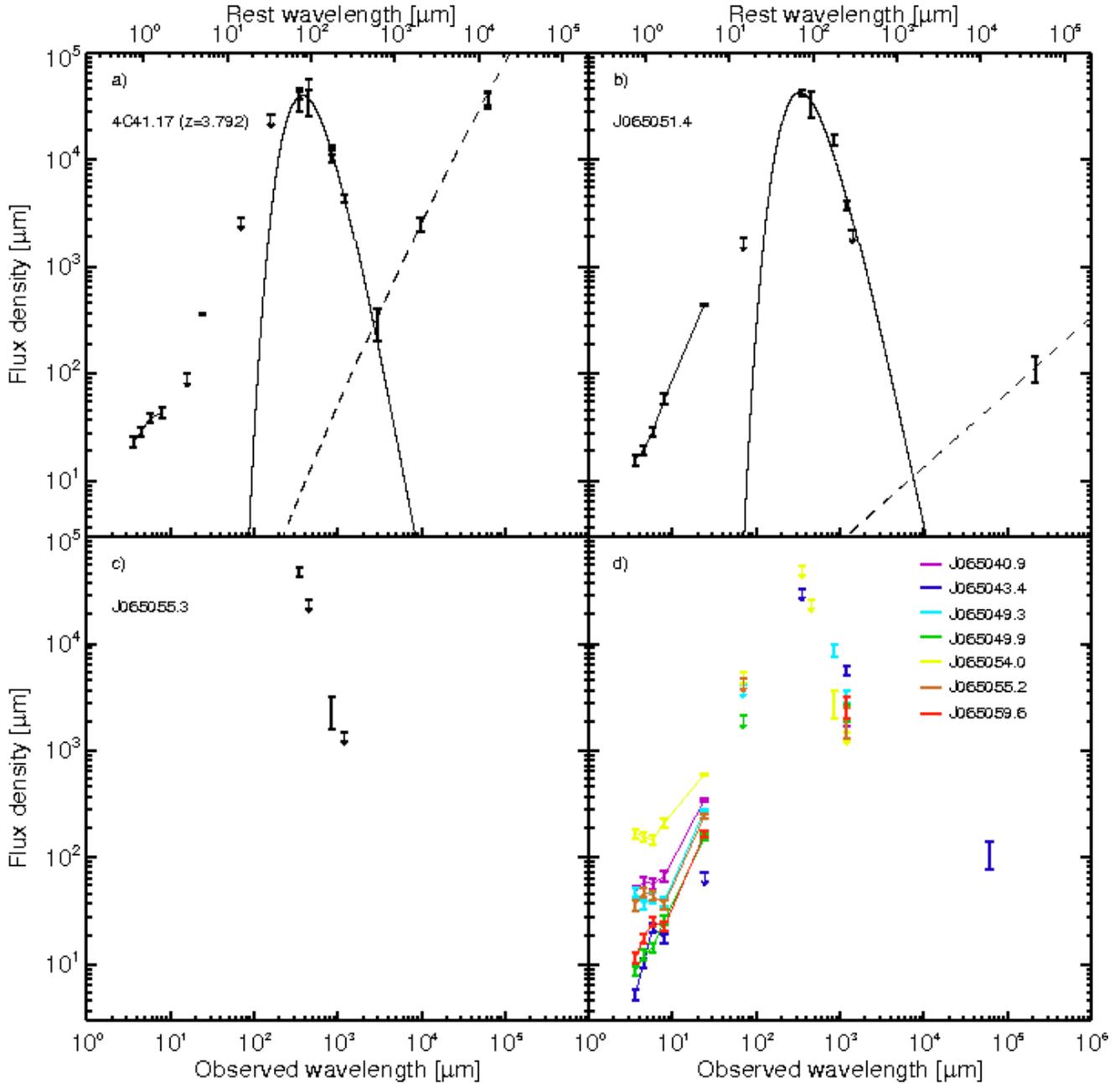}
\caption[]{ {\bf a)-c)} Radio/mm/mid-IR SEDs of 4C\,41.17 and the two
SMGs detected at 350-$\mu$m, J065051.4 and J065055.3. Notice that in
the case of 4C\,41.17 we have included the 16-$\mu$m upper flux limit
of $3\sigma \ls 99\,\mu$Jy from Seymour et al. (2007). In the case of
4C\,41.17 and J065051.4 greybody laws were fitted to the far-IR/submm
data while spline fits were used to model the mid-IR part of the
spectra. The dust temperature ($T_d$), spectral index ($\beta$) and an
overall normalization constant were allowed to vary freely in the
fitting process. The radio spectrum of 4C\,41.17 was fitted with a
parabola, while a power-law with slope $\alpha=-0.7$ (dashed curve),
consistent with the radio measurement, was adopted for J065051.4.
{\bf d)} Radio/mm/mid-IR SEDs of the remaining 7 SMGs with robust
radio and/or 24-$\mu$m counterparts.  }
\label{figure:seds}
\end{center}
\end{figure*}

The mid-IR SEDs of all eight SMGs (and 4C\,41.17) identified in the
radio and/or the mid-IR are shown in Fig.\ \ref{figure:seds}. In the
case of 4C\,41.17 the IRAC measurements clearly deviate from a
straight power-law, and instead seem to resemble a bump indicative of
a stellar population. At $z=3.792$, however, the 8.0-$\mu$m IRAC band
samples the rest-frame SED at 1.7-$\mu$m, i.e.\ just longward of
1.6-$\mu$m, thus making it hard to discern the exact shape of the
stellar bump.  Moreover, the 8.0-$\mu$m flux is likely to be affected
by an extremely hot, AGN-heated dust component (several hundreds
Kelvin hot), the presence of which is suggested by the 24-$\mu$m
flux. While the lack of 16-, 70- and 160-$\mu$m detections of
4C\,41.17 (Seymour et al.\ 2007) makes it impossible to constrain the
temperature of this hot dust component, the 3-$\sigma$ upper limits
that we are able to derive are clearly consistent with the presence of
a hot dust component.

The mid-IR spectrum of J065051.4 is consistent with a power law
through all of the IRAC bands, suggesting the stellar emission is
swamped by the emission from the hot, AGN-heated dust. Thus, the near-
and mid-IR properties of this source indicate the presence of an AGN
more strongly than in 4C\,41.17, yet J065051.4 is more than two orders
of magnitude less radio-luminous than 4C\,41.17. There is a large
spread in the observed radio-loudness of AGN, and only $\sim 10$ per
cent of them are radio-loud (e.g.\ Visnovsky et al.\ 1992; Hooper et
al.\ 1995; Stern et al.\ 2000), making it more likely to find a
radio-quiet AGN in the field of a radio galaxy than to find another
radio-loud AGN. In fact several examples of such AGN, detected in the
fields of radio galaxies, have been reported (Pentericci et al.\ 2000;
Venemans et al.\ 2007).

All of the remaining seven SMGs with IRAC and/or MIPS detections
(Fig.\ \ref{figure:seds}d) show some evidence of a stellar
population. Some exhibit a clear stellar bump (J065049.9, J065043.4,
J065055.2 and J065059.6), others only a hint thereof (J065049.3,
J065049.9 and J065054.0). In fact J065049.9 appears to be a mixed
AGN/starburst system with a mid-IR spectrum similar to that of
4C\,41.17.

\begin{figure}
\begin{center}
\includegraphics[width=1.0\hsize,angle=0]{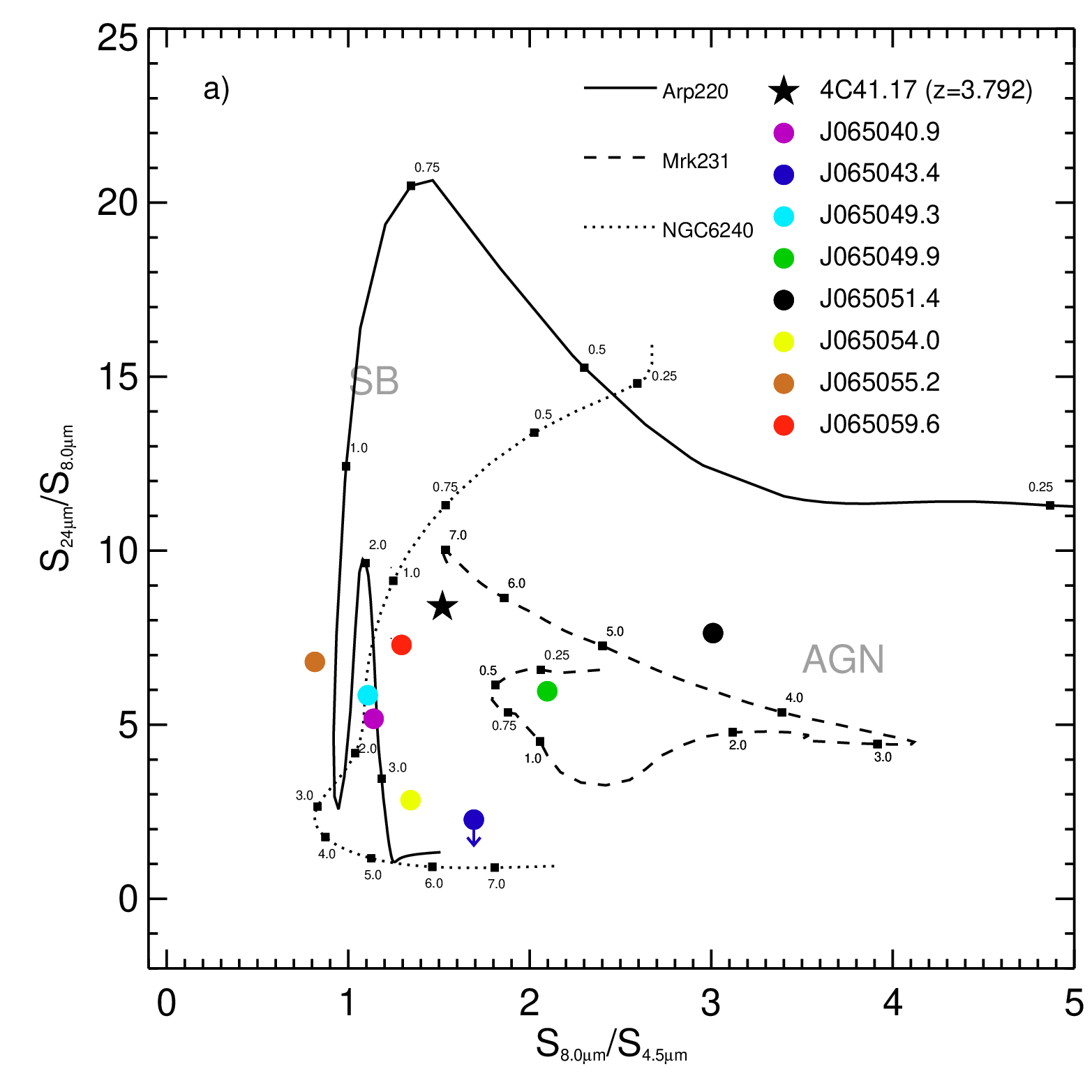}
\includegraphics[width=1.0\hsize,angle=0]{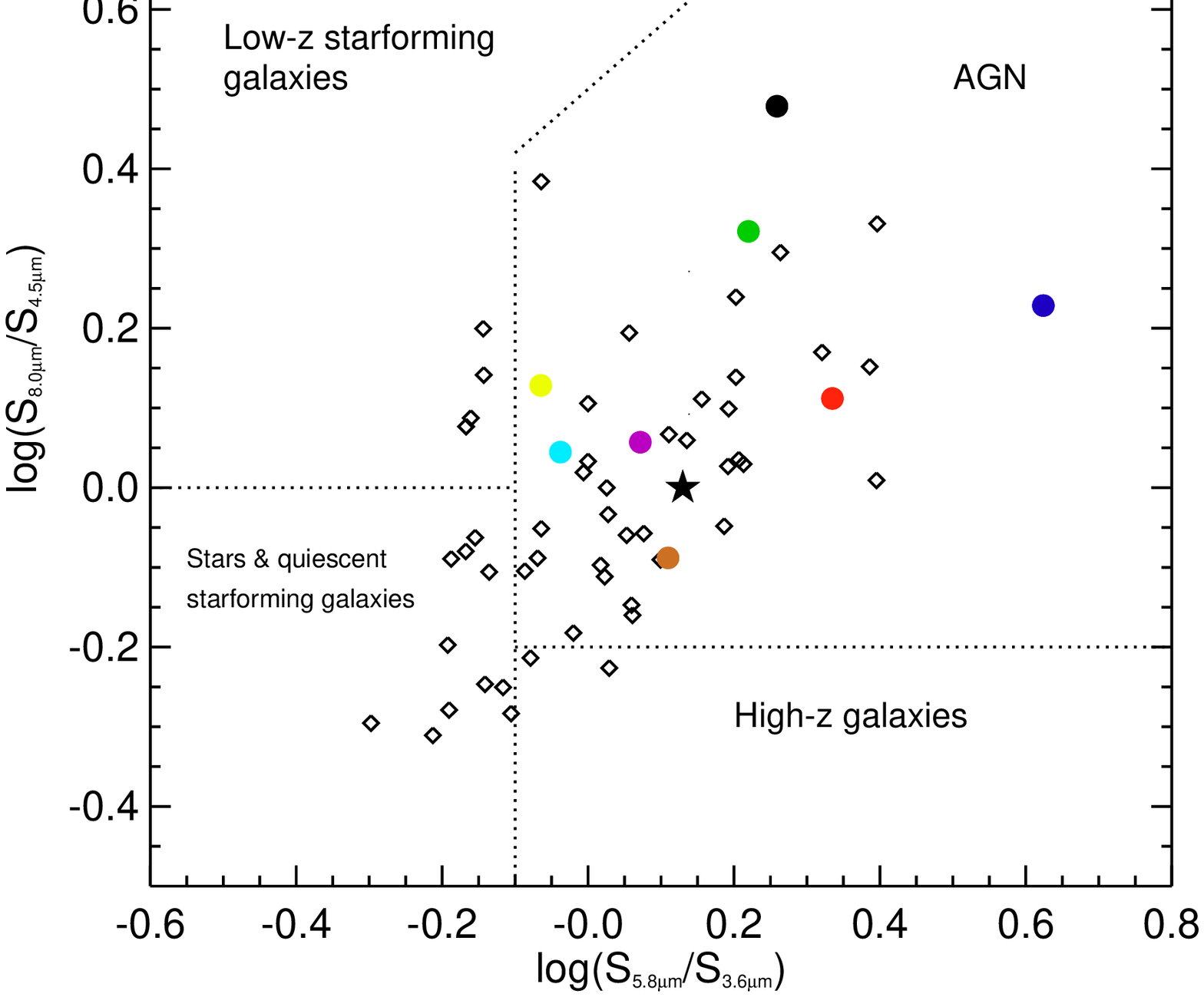}
\caption[]{{\bf a)} The $S_{24\mu m}/S_{8.0\mu m}$ vs.\ $S_{8.0\mu
m}/S_{4.5\mu m}$ {\it Spitzer} colour-colour diagram for 4C\,41.17 and
the 8 SMGs within the MAMBO map with robust mid-IR counterparts,
identified either via the mid-IR or the radio.  The curves track the
location of the local (Ultra) Luminous Infra-Red Galaxies Arp\,220
(solid curve), NGC\,6240 (dotted line), and Mrk231 (dashed line) in
the diagram as a function of redshift. {\bf b)} The $S_{8.0\mu
m}/S_{4.5\mu m}$ vs.\ $S_{5.8\mu m}/S_{3.6\mu m}$ {\it Spitzer}
colour-colour diagram (Stern et al.\ 2005). The colour doing is the
same as in a). The diamonds represent a sample of blank-field SMGs
from the SHADES survey (Ivison et al.\ 2007).  }
\label{figure:mir}
\end{center}
\end{figure}

This picture is confirmed in Fig.\ \ref{figure:mir}a where we have
plotted $S_{24\mu m}/S_{8.0\mu m}$ vs.\ $S_{8.0\mu m}/S_{4.5\mu m}$
for 4C\,41.17 and the eight SMGs with robust mid-IR counterparts. This
{\it Spitzer} colour-colour diagram can serve as a discriminator
between pure AGN-dominated systems and pure starbursts (Ivison et al.\
2004, 2007), as illustrated by the clear separation between the
Arp\,220 (starburst) and Mrk231 (AGN) tracks in the diagram.
NGC\,6240, which is thought to be a 'mixed' system with significant
contributions to its mid-IR/far-IR luminosity from both a starburst
and an AGN lies in the intermediate region. Clearly, J065051.4 and
J065049.9 occupy the AGN-region of the diagram, while the remaining
seven SMGs fall within starburst-region outlined by the Arp220 track.
4C\,41.17 itself falls midway the two regions, close to the NGC\,6240
track, consistent with its spectra showing evidence of a stellar
population as well as an AGN.

Lacy et al.\ (2004) proposed to use the $\log S_{5.8\mu m}/S_{3.6\mu
m}$ vs.\ $\log S_{8.0\mu m}/S_{4.5\mu m}$ as a means to identify
obscured AGN as well as unobscured AGNs (see also Stern et al.\
2005). Fig.\ \ref{figure:mir}b shows that our sources along with a
sample of blank-field SMGs (Ivison et al.\ 2007) largely fall within
the region selecting AGN. As much as four (or 50 per cent) of our
sources (J065043.4, J065049.9, J065051.4 and J065059.6) are very red
in both colours, and therefore likely to be AGN dominated
systems. One of the these sources is J065043.4, which we show in
\S \ref{subsection:mambo-blob} is a highly extended Ly$\alpha$ blob at
$z=2.672$. Two of the sources (J065049.9 and J065051.4) with very
red colours have photometric redshifts which put them at the same
redshifts as 4C\,41.17, while the fourth source (J065059.6) is
estimated to lie at $z\sim 2.6$.

The remaining four sources, namely J065040.9, J065054.0, J065049.3 and
J065055.2, are closer to the boundaries where contamination by
star-forming galaxies becomes important (Stern al.\ 2005).  This is
broadly consistent with the AGN vs.\ starburst classification we
derived from the colour-colour diagram in Fig.\ \ref{figure:mir}a,
where these four sources lie in the starburst part of the diagram.

Finally, we note that while half of our SMG sample have very red
$S_{5.8\mu m}/S_{3.6\mu m}$ and $S_{8.0\mu m}/S_{4.5\mu m}$ colours,
and therefore likely to be AGN dominated systems, the fraction of SMGs
selected from blank-field surveys with similarly red colours is much
smaller ($\sim 7$ per cent). Thus the fraction of AGN dominated system
seems to be higher in our sample than in the general SMG population.
It is possible that the fact that we find two likely AGN dominated
systems (J065049.9 and J065051.4) at the same redshift as the radio
galaxy is evidence for an excess of AGNs in the proto-cluster
associated with 4C\,41.17 -- much like what has been seen in more
nearby clusters (Johnson, Best \& Almaini 2003; Eckart et al.\ 2006).

%
%
\begin{table}
\scriptsize
\caption{IRAC fluxes of the MIPS and radio counterparts given in Table \ref{table:mips-radio-fluxes}. The fluxes
were extracted from a 9.26\arcsecs~aperture using aperture corrections of 1.091(3.6-$\mu$m), 1.117(4.5-$\mu$m), 1.100(5.8-$\mu$m),
and 1.165(8.0-$\mu$m) from Lacy et al.\ (2005).} 
\begin{center}
\begin{tabular}{lccccccccccc}
\hline
\hline
ID                 &  $S_{3.6\mu\rm{m}}$   & $S_{4.5\mu\rm{m}}$    & $S_{5.8\mu\rm{m}}$    & $S_{8.0\mu\rm{m}}$    \\
                   &                       &                       &                       &                       \\ 
                   & ($\mu$Jy)             & ($\mu$Jy)             &  ($\mu$Jy)            & ($\mu$Jy)             \\ \hline
4C\,41.17          & $22.9\pm 2.4$         & $28.7 \pm 3.0$        & $38.7\pm 3.9$         & $43.6\pm 4.4$         \\    
J065040.0          & $118.3\pm 11.9$       & $89.5 \pm 9.0$        &  $67.2\pm 6.8$        & $43.4\pm 4.4$         \\    
J065040.9          & $48.5\pm 4.9$         & $59.2 \pm 6.0$        &  $57.2\pm 5.8$        & $67.5\pm 6.8$         \\    
MIPS-1             & $68.3\pm 6.9$         & $57.2 \pm 5.8$        &  $60.7\pm 6.1$        & $53.2\pm 5.3$         \\    
J065043.4          & $5.3\pm 0.6$          & $10.4 \pm 1.1$        &  $22.3\pm 2.3$        & $17.6\pm 1.8$         \\    
                   & $18.1\pm 1.9$         & $23.4 \pm 2.4$        &  $30.1\pm 3.1$        & $14.0\pm 1.4$         \\    
                   & $16.9\pm 1.8$         & $18.4 \pm 1.9$        &  $23.0\pm 2.3$        & $11.2\pm 1.2$         \\    
J065047.5          & $38.5\pm 3.9$         & $36.1 \pm 3.7$        &  $34.1\pm 3.5$        & $25.2\pm 2.6$         \\    
J065049.3          & $39.5\pm 4.0$         & $42.6 \pm 4.4$        &  $36.2\pm 3.7$        & $47.2\pm 4.8$         \\    
                   & $21.7\pm 2.3$         & $29.4 \pm 3.0$        & $30.8\pm 3.1$         & $26.4\pm 2.7$         \\    
                   & $12.2\pm 1.3$         & $16.3 \pm 1.7$        & $20.8\pm 2.1$         & $19.3\pm 2.0$         \\    
                   & $10.1\pm 1.1$         & $12.0 \pm 1.3$        & $18.5\pm 1.9$         & $13.7\pm 1.4$         \\    
J065049.9          & $8.8\pm 1.0$          & $12.4 \pm 1.3$        &  $14.6\pm 1.5$        & $26.0\pm 2.6$         \\    
J065051.4          & $15.7\pm 1.7$         & $19.5 \pm 2.0$        &  $28.5\pm 2.9$        & $58.7\pm 5.9$         \\    
J065052.5          & $42.6\pm 4.3$         & $56.0 \pm 5.7$        &  $47.1\pm 4.8$        & $43.7\pm 4.4$         \\    
J065054.0          & $170.4\pm 17.1$       & $158.7 \pm 16.0$      &  $146.8\pm 14.7$      & $213.2\pm 21.4$       \\    
J065055.2          & $36.9\pm 3.8$         & $44.1 \pm 4.5$        &  $47.5\pm 4.8$        & $36.0\pm 4.6$         \\    
                   & $21.7\pm 2.3$         & $28.7 \pm 3.0$        & $29.9\pm 3.0$         & $21.9\pm 2.2$         \\    
J065055.3          & $55.2\pm 5.6$         & $45.0 \pm 4.6$        &  $38.8\pm 3.9$        & $38.1\pm 3.8$         \\    
J065059.3          & $3\sigma \le 1.0$     & $1.8 \pm 1.0$         &  $9.9\pm 1.0$         & $13.9\pm 6.5$         \\    
J065059.6          & $11.7\pm 1.3$         & $17.7 \pm 1.9$        &  $25.3\pm 2.6$        & $22.9\pm 2.3$         \\    
                   & $15.5\pm 1.6$         & $11.4 \pm 1.2$        & $8.2\pm 0.9$          & $4.9\pm 0.6$          \\    
                   & $49.5\pm 5.0$         & $59.3 \pm 6.0$        &  $64.6\pm 6.5$        & $78.5\pm 7.9$         \\    
J065104.0          & . . .                 & $5.8 \pm 2.0$         &  . . .                & $\le 10.1$            \\    
                   & . . .                 & $6.6 \pm 2.0$         &  . . .                & $13.9\pm 7.3$         \\    
\hline
\label{table:irac-fluxes}
\end{tabular}
\end{center}
\end{table}
%
%

\subsection{Photometric redshifts}\label{section:phot-z}
In addition to serving as a crude starburst vs.\ AGN diagnostic, the
1.6-$\mu$m stellar 'bump' may also serve as a means of obtaining
photometric redshifts (e.g.\ Sawicki 2002). Even though photometric
redshifts derived in this way are bound to be uncertain, they might be
helpful for our primary purpose of establishing or refuting an
over-density of SMGs associated with 4C\,41.17 by simply being able to
conclude whether a source is at a redshift $< 4$ or $> 4$.

We therefore proceeded to derive photometric redshifts for the eight
SMGs (and 4C\,41.17) with robust mid-IR/radio identifications.  The
photometric redshifts were estimated simply by identifying the IRAC
channel most likely to represent the 1.6-$\mu$m rest-frame stellar
feature. The results are listed in Table \ref{table:redshifts}. In
addition to the 8 SMGs and 4C\,41.17, we also included a
spectroscopically identified 24-$\mu$m source in order to see how well
our photometric redshift technique recovers the true redshift (this
source is located 11.6\arcsecs~ south-west of J065040.9 and is denoted
MIPS-1 -- Table \ref{table:mips-radio-fluxes}).

%
%
\begin{table}
\scriptsize
\caption{Spectroscopic and photometric redshifts.} 
\begin{center}
\begin{tabular}{lccc}
\hline
\hline
ID                 &  $z_{\rm phot,IRAC}$   & $z_{\rm phot,radio/submm}$  & $z_{\rm spec}$            \\
\hline
4C\,41.17          &  $\sim 4$              &                         & $3.792\pm 0.001$      \\ 
J065040.0          &  . . .                 & . . .                   & . . .                 \\ 
J065040.9          &  $\sim 1.8$            & . . .                   & . . .                 \\ 
 MIPS-1            &  $< 1.3$               & . . .                   & $0.909\pm 0.002$      \\ 
J065043.4          &  $\sim 2.6$            & $> 2.8$                 & $2.672\pm 0.001$      \\ 
J065047.5          &  . . .                 & $> 1.7$                 & . . .                 \\ 
J065049.3          &  $< 1.3$               & $1.2-16$                & $1.184\pm 0.002$      \\ 
J065049.9          &  $\sim 4$              & . . .                   & . . .                 \\ 
J065051.4          &  . . .                 & $2.0-4.3$               & . . .                 \\ 
J065052.5          &  . . .                 & . . .                   & . . .                 \\ 
J065054.0          &  $< 1.3$               & $1.5-16$                & $0.507\pm0.020$       \\ 
J065055.2          &  $\sim 1.8$            & . . .                   & . . .                 \\ 
J065055.3          &  . . .                 & . . .                   & . . .                 \\ 
J065059.3          &  . . .                 & . . .                   & . . .                 \\ 
J065059.6          &  $\sim 2.6$            & . . .                   & . . .                 \\ 
J065104.0          &  . . .                 & . . .                   & . . .                 \\ 
\hline
\label{table:redshifts}
\end{tabular}
\end{center}
\vspace*{-0.5cm}
\end{table}
For the five sources (4C\,41.17, J065040.9, J065043.4, J065055.2 and
J065059.6) which exhibit a relatively well defined stellar 'bump' we
are able to infer a single value for the photometric redshifts, while
only upper limits can be inferred for the remaining three sources
(J065049.3, J065054.0 and MIPS-1) where the stellar bump appears to
have been redshifted into the 3.6-$\mu$m channel or blueward of
it. The upper limits derived in those three latter cases assume that
the 1.6-$\mu$m (rest-frame) stellar feature falls in or below the
3.6-$\mu$m (observed frame) channel.

Only two of the five sources with spectroscopic redshifts have
well-defined stellar bumps, but in both of those cases (4C\,41.17 and
J065043.4) we find good agreement between their photometric and
spectroscopic redshifts.  The upper limits on the photometric
redshifts for the remaining three sources (J065049.3, J065054.0 and
MIPS-1) are consistent with their spectroscopic redshifts. Thus it
seems that not only are redshift estimates based on the 1.6-$\mu$m
feature robust when a well-defined stellar bump is present in the
mid-IR spectrum, but that meaningful upper limits can be placed even
when this feature is absent (and the spectrum is not dominated by an
AGN).  The photometric redshifts for J065040.9, J065055.2 and
J065059.6 are $\sim 1.8$, $\sim 1.8$ and $\sim 2.6$, respectively, but
unfortunately we do not have spectroscopic redshifts for any of those
three sources.
%
%
%
%
\begin{figure}
\begin{center}
\includegraphics[width=0.9\hsize,angle=0]{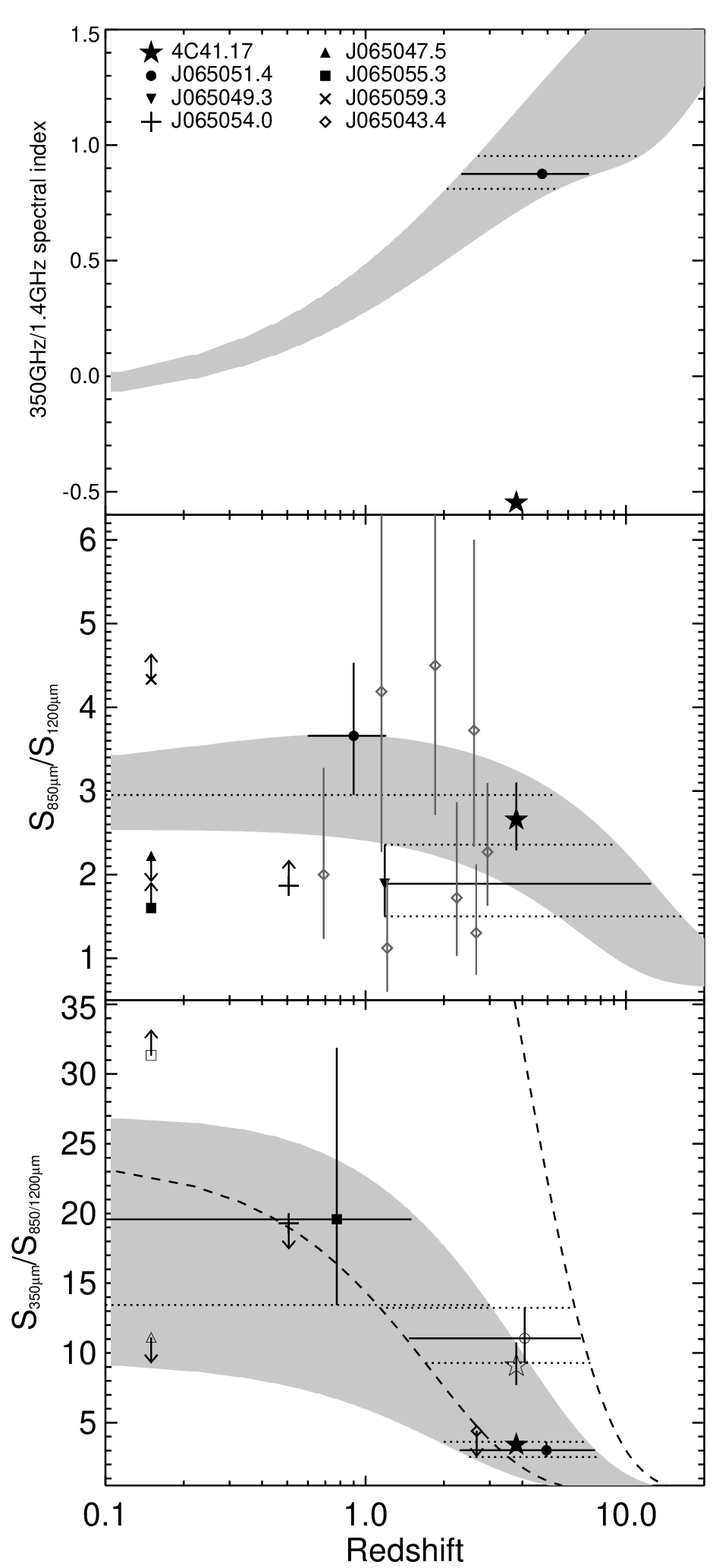}
\caption[]{{\bf Top:} The 1.4GHz-to-350GHz spectral indices for 4C\,41.17 and J05651.4. 
{\bf Middle:} The 850-/1200-$\mu$m flux ratio
for 4C\,41.17 and the 6 SMGs in our sample detected with SCUBA and/or
MAMBO (Table \ref{table:data}). Also shown are the flux ratios for a
sample of robust blank-field SMGs identified from SCUBA and MAMBO
surveys of the Lockman Hole East (Greve et al.\ 2004b; Ivison et al.\
2005).  {\bf Bottom:} The 350-/850-$\mu$m (filled symbols) and
350-/1200-$\mu$m (open symbols) for the SMGs in our sample which have
been observed at 350-$\mu$m and 850- or 1200-$\mu$m. In two top panels
the grey shaded region outlines the range of flux ratios allowed by
modified black body SEDs with $T_d = 20-70\,$K and $\beta =
1.0-2.0$. In the lower panel, the grey-shaded region and the region
outlined by the dashed curves mark the range of 350-/850-$\mu$m flux
and 350-/1200-$\mu$m flux ratios, respectively, as a function of
redshift allowed by these models. Where possible, the sources have
been placed at their spectroscopic redshifts. Otherwise, sources were
placed at the center of the photometric redshift range suggested by
their flux ratio, or, in the case where only upper/lower limits were
available, placed at $z=0.15$.  For sources with well determined flux
ratios, horizontal dotted lines have been drawn to outline the range
of possible redshifts allowed by the SEDs models within the errors of
the flux ratios.  }
\label{figure:z_submm}
\end{center}
\end{figure}

\bigskip

In a further attempt to get a handle on the redshifts of our SMG
sample we have looked at their radio/submm/mm colours as photometric
redshift indicators.  Historically, the first photometric redshift
indicator considered for SMGs was the 850-$\mu$m/1.4-GHz spectral
index (Hughes et al.\ 1998; Carilli \& Yun 1999), which makes use of
the steep opposite slopes of the radio and (sub)mm parts of a typical
starburst SED.  The radio-submm spectral index, however, is only
useful as a redshift indicator at redshifts $\ls 3$, due partly to the
rapid drop-off in the detectability of SMGs at higher redshifts in
even the deepest radio maps. Of the (sub)mm colours one may use the
850-/1200-$\mu$m flux ratio, for example, which is expected to be a
strong function of redshift beyond $z\gs3$ (Eales et al.\ 2003; Greve
et al.\ 2004b), while the 350-/850-$\mu$m and 350-/1200-$\mu$m flux
ratios are mostly sensitive to the redshift range $1\ls z \ls 3$.  In
Fig.\ \ref{figure:z_submm} we compare the above flux ratios as derived
for our sample with the predicted flux ratios (as a function of
redshift) for a broad range of possible greybody far-IR/submm SEDs
($T_d = 35-65\,$K and $\beta = 1.0-2.0$), and use these model
predictions to constrain the possible redshift range for our sources.

As expected given that 4C\,41.17 is a radio-loud AGN, its
850-$\mu$m-to-radio spectral index ($\alpha = 0.42\log (S_{\rm
1.4GHz}/S_{\rm 350GHz})$) lies below the model predictions by more
than an order of magnitude. Its (sub)mm flux ratios, on the other
hand, are consistent with the SED models at $z\simeq 3.8$.

The radio-to-submm index of J065051.4 places it at a redshift of
$2.1\ls z \ls 7.2$. The 850-/1200-$\mu$m flux ratio of J065051.4 is at
the high end of what is typically observed in blank-field SMGs but
consistent with the models, given the large error bars. We put an
upper limit on the redshift of this source of $z\ls 5.2$.  The low
350-/850-$\mu$m and 350-/1200-$\mu$m flux ratios of this source yield
possible redshift ranges of $2.0\ls z \ls 7.7$ and $1.2\ls z \ls 4.3$,
respectively.  Thus, the four flux ratios for J065051.4 yield
consistent photometric redshift ranges, and assuming that the true
redshift should be found in the overlap region of these four
redshifts-ranges, we constrain the likely redshift range for this
source to $2.0\ls z \ls 4.3$.

The 850-/1200-$\mu$m flux ratio for J065049.3 is on the low side,
barely consistent within the errors, of the expected range of flux
ratios predicted by the model and thus, in the absence of a
spectroscopic redshift, would favour a redshift range of $1.2\gs z\gs
16$, where the uncertainties on the flux ratio have been taken
into account. The low-end of this redshift range is consistent with
the spectroscopic redshift ($z=1.184$), and suggest that we have
correctly identified the mid-IR counterpart to this SMG and thus its
redshift.  This scenario is further strengthened by the X-ray and
$z$-band detections of the favoured 24-$\mu$m candidate
counterpart. Also, we note that the flux ratio is well within the
range observed for blank-field SMGs with spectroscopic redshifts in
the range $0.7\ls z\ls 3$.

For the remaining sources, only upper and/or lower limits are
available for the various flux ratios, severely weakening the
constraints that can be put on the redshifts of these sources. In the
case of J065054.0 where the redshift is known from spectroscopy, we
find the upper limit on the 350-/850-$\mu$m flux ratio for this source
is consistent with the model values at $z=0.507$. For J065043.4 where
we find a robust spectroscopic redshift of $z=2.672$ we find a
350-/1200-$\mu$m flux ratio which is consistent (albeit marginally)
with the range of flux ratios from the models.  Of the three sources
with no spectroscopic redshifts we find that J065047.5 has consistent
350-/1200-$\mu$m and 850-/1200-$\mu$m flux ratios which both point
towards high redshifts ($z\gs 1.7$).

\subsection{Far-IR luminosities, dust temperatures and masses}\label{subsection:lfir}
SHARC-II observations provide a particularly important datum in the
SEDs of galaxies at redshifts beyond unity. At such redshifts,
350-$\mu$m observations sample very close to the rest-frame dust peak
at $\sim$100-$\mu$m, allowing for more accurate estimates of dust
temperature and far-IR luminosity than can be made with 850- and/or
1200-$\mu$m observations only. SCUBA and MAMBO observations, however,
are better at constraining the Rayleigh-Jeans tail, and the
combination of short and long (submm) wavelengths observations is
therefore potentially very powerful, allowing for a much better
characterization of the full far-IR/submm SED.

Two of the SMGs (J065051.4 and J065055.3) presented here, and
4C\,41.17 itself, were detected at 350-$\mu$m in addition to 850- and/or 
1200-$\mu$m. The radio and far-IR/mm SEDs of these three sources are shown 
in Fig.\ \ref{figure:seds} along with the remaining five SMGs with 
robust 24-$\mu$m/1.4-GHz identifications (see Table \ref{table:data}).

Only 4C\,41.17 and J065051.4 have robust detections at four or more
(sub)mm wavelengths, and as a result we only attempted to derive dust
temperatures and spectral indices for those two sources.  The SEDs
were fit with an optically thin greybody law of the form: $S_{\nu_o}
\propto \nu_r^{\beta} B(\nu_r, T_d)$, where $S_{\nu_o}$ is the flux at
the observed frequency $\nu_o$, $\beta$ is the grain emissivity, $T_d$
the dust temperature, and $B(\nu,T_d)$ the Planck function.  All the
far-IR/submm data listed in Table \ref{table:data} for these two
sources, except for upper limits, were used in the fit. All three
parameters involved in the fit, namely $T_d$, $\beta$ and an overall
normalization factor, were allowed to vary freely.  Also, given the
high likelihood that J065051.4 is at the same redshift as 4C\,41.17
(\S \ref{subsection:overdense}), we adopted a redshift of $z=3.792$
when fitting the SED of this source. The resulting greybody fits are
shown as solid curves in Fig.\ \ref{figure:seds}a and b. The
corresponding physical quantities, including the far-IR luminosity,
dust masses and star-formation rate, are listed in Table
\ref{table:results}.

We also include the radio data in Fig.\ \ref{figure:seds} in order to
compare with the far-IR/submm part of the SED, and in particular to
check whether the synchrotron radiation contributes significantly to
the thermal dust emission in any of the sources. For 4C\,41.17 radio
data was compiled from the WENSS (325\,MHz; Rengelink et al.\ 1997),
Texas (365\,MHz; Douglas et al.\ 1996), Green Bank Northern Sky Survey
(1.4\,GHz; White \& Becker 1992) and the NRAO VLA Sky Survey
(1.4\,GHz; Condon et al.\ 1998), and fitted with a parabolic
function. While the synchrotron emission can account for all of the
observed 3\,mm emission, its contribution to the emission at (sub)mm
wavelengths is negligible ($\ls 1$ per cent of the measured 1200- or
850-$\mu$m flux densities). Thus, as expected, the (sub)mm emission
originating from 4C\,41.17 is dominated by thermal emission from dust
(Dunlop et al.\ 1994). The same goes for J065051.4, if we assume
a radio spectrum normalized to the 1.4-GHz flux and with a slope of
$\alpha = -0.7$ (where $S_{\nu} \propto \nu^{\alpha}$).

For the sources for which spectroscopic or reliable photometric
redshifts were available, far-IR luminosities were derived by
integrating their SEDs over the wavelength range $\lambda = 40 -
1000\,\mu$m. While we were able to integrate over the best-fit SED in
the case of 4C\,41.17 and J065051.4, for the remaining seven sources
with redshift estimates we simply adopted a greybody SED (with
$T_d=40\,$K and $\beta=1.5$), normalized to its 850-$\mu$m flux.  When
no 850-$\mu$m flux was available, we used 1200-$\mu$m flux densities
(or if that wasn't available, the 350-$\mu$m flux density) together
with the scaling relation, $S_{\nu} \propto \nu^{2+\beta}$ to derive
$S_{850\mu m}$.  Far-IR luminosities for the last 6 sources with no
robust redshift estimate were derived using
$L_{\mbox{\tiny{FIR}}}/\Lsolar = 1.9\times 10^{12}\,(S_{850\mu m}/\rm
mJy)$ which is valid for a greybody SED with $T_d = 40\,$K and $\beta
= 1.5$.  The above relation for the far-IR luminosity applies over the
redshift range $\sim2-4$ with typical uncertainties of about a factor
of $\sim 2-3$ (Blain et al. 2002); but the uncertainties could of
course be even higher if the sources do not lie in this redshift
range.

The dust masses giving rise to the far-IR/submm emission were estimated 
following the prescription by Hildebrand (1983): 
\begin{equation} M_d =
\frac{1}{1+z}\frac{S_{\nu_o}D^2_L}{\kappa(\nu_r) B(\nu_r, T_d)},
\end{equation} 
where $\nu_o$ and $\nu_r$ are the observed and rest-frame frequencies,
respectively, $S_{\nu_o}$ is the observed flux density, $D_L$ is the
luminosity distance, and $\kappa(\nu_r)$ is the mass absorption
coefficient in the rest frame. For the latter we have adopted a value
of $\kappa(\nu_r) = 0.11 (\nu_r / 352\,\mbox{GHz})^{\beta}$ in units
of m$^2$\,kg$^{-1}$. Dust masses were only derived for sources with
redshift estimates. Star-formation rates were estimated using
$SFR/\Msolar\,\mbox{yr}^{-1} = L_{\mbox{\tiny{FIR}}}/1.7\times
10^{10}\,\Lsolar$ (Kennicutt et al.\ 1998).  The resulting far-IR
luminosities, dust masses and star-formation rates are listed in Table
\ref{table:results}. The star-formation rates should be
considered as upper limits as they were derived on the basis that the
far-IR luminosities of these systems are dominated by star-formation,
and not AGN-activity.

%
%
\begin{table*}
\caption{Derived physical properties for 4C\,41.17 and the 14 SMGs in our sample. Redshifts in
parentheses are photometric redshifts.}
\vspace{0.2cm}
\begin{center}
\begin{tabular}{lcccccc}\hline
\hline
Source           &      & $T_d$      & $\beta$      & $L_{\mbox{\tiny{FIR}}}$    &  $M_{d}$                  & $SFR$                   \\ 
                 &      & [K]        &              & [$\times 10^{13} \Lsolar$] &  [$\times 10^9 \Msolar$]  & [$\Msolar\,\mbox{yr}^{-1}$]\\
\hline
4C\,41.17        & 3.792& $44\pm 2$  & $1.6\pm 0.1$ & $1.5$                      & $0.8$                     & $900$\\
J065051.4        & 3.792& $47\pm 3$  & $1.6\pm 0.1$ & $2.1$                      & $0.8$                     & $1200$\\
\hline
J065040.0$\dagger$& . . .&  40.0      & 1.5          & $2.2$                      & $. . .$                   & $1300$\\
J065040.9        & (1.8) &  . . .     & . . .        & $0.9$                      & $1.0$                     & $500$\\
J065043.4        & 2.672 &  . . .     & . . .        & $2.3$                      & $2.5$                     & $1300$\\
J065047.5$\dagger$& . . .&  . . .     & . . .        & $1.7$                      & $. . .$                   & $1000$\\
J065049.3        & 1.184 &  . . .     & . . .        & $0.6$                      & $0.7$                     & $400$\\
J065049.9        & (4)   &  . . .     & . . .        & $0.9$                      & $1.0$                     & $500$\\
J065052.5$\dagger$& . . .&  . . .     & . . .        & $2.0$                      & $. . .$                   & $1200$\\
J065054.0        & 0.507 &  . . .     & . . .        & $0.1$                      & $0.1$                     & $100$\\
J065055.2        & (1.8) &  . . .     & . . .        & $0.8$                      & $0.9$                     & $500$\\
J065055.3$\dagger$& . . .&  . . .     & . . .        & $0.5$                      & $. . .$                   & $300$\\
J065059.3$\dagger$& . . .&  . . .     & . . .        & $1.2$                      & $. . .$                   & $700$\\
J065059.6        & (2.6) &  . . .     & . . .        & $1.1$                      & $1.2$                     & $600$\\
J065104.0$\dagger$& . . .&  . . .     & . . .        & $2.7$                      & $. . .$                   & $1600$\\
\hline
\end{tabular}
\end{center}
$\dagger$ The far-IR luminosity was calculated using $L_{\mbox{\tiny{FIR}}}/\Lsolar = 1.9\times 10^{12}\,(S_{850\mu m}/\rm mJy)$, see
\S\ref{subsection:lfir} for details. 
\label{table:results}
\vspace*{-0.2cm}
\end{table*}

\section{Discussion}\label{section:discussion}

\subsection{Is 4C\,41.17 a highly over-dense region of the Universe?}\label{subsection:overdense}
Taken at face value, the 14 SMGs (not including 4C\,41.17 itself)
detected within the 58 sq.\ arcmin region outlined by the MAMBO map
corresponds to a surface density of $\sim 0.24\,$sq.\ arcmin$^{-2}$
or, equivalently, $860\,$sq.\ deg$^{-2}$. If we adopt a more
conservative approach and only include those 11 SMGs which have either
been confirmed or detected for the first time at $\ge
3.5$-$\sigma$ at 1200-$\mu$m with MAMBO we find an average surface
density of $\sim 675\,$sq.\ deg$^{-2}$.

This is perfectly consistent with the 1200-$\mu$m blank-field number
counts which yield $N(S_{1200\mu m}\ge 2\,\mbox{mJy}) \sim
700\,$deg$^{-2}$ -- so obtained by (mildly) extrapolating from the
observed 1200-$\mu$m counts in the flux range $\sim 2.8 - 5.3\,$mJy
(Greve et al.\ 2004b).  Thus, even in the most optimistic scenario
where all 14 SMGs are real, we find the surface density of SMGs, when
averaged over the entire MAMBO map, to be no more than $\sim 20\,$
per-cent above the blank field.  It is possible, however, that by
averaging over the entire MAMBO map, which is a relatively large area,
any presence of a SMG-over-density may have been 'diluted', and thus
does not appear to be significant. If 4C\,41.17 really is an over-dense
region, however, we would expect a steady increase in the source
concentration towards the radio galaxy, with most of the MAMBO sources
at the outskirts of the map belonging to the blank field SMG
population, while the fraction of SMGs associated with 4C\,41.17
increases towards the center.
%
%
\begin{figure}
\begin{center}
\includegraphics[width=1.0\hsize,angle=0]{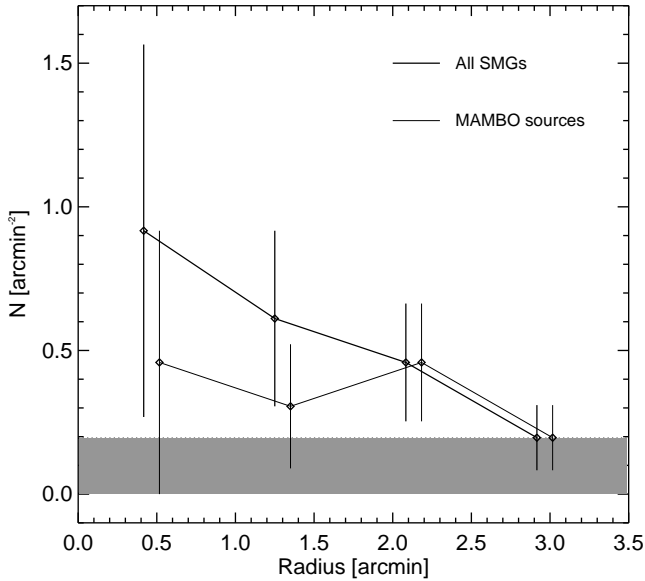}
\caption[]{The average surface density of SMGs within 50\arcsecs~wide
annuli at 50, 100, 150, and 200\arcsecs~from 4C\,41.17 (but not
including the radio galaxy itself).  The thick curve represents the
surface density profile using all 14 SMGs in our sample, while the
thin curve is derived using only the sources detected at 1200-$\mu$m
with MAMBO. The error bars are Poisson errors. The grey shaded region
outlines the blank field surface density of 1200-$\mu$m selected
sources brighter than $2\,$mJy ($\sim 0.19\,$sq.\ arcmin.$^{-1}$ --
upper edge) and $10\,$mJy ($\sim 0.03\,$sq.\ arcmin.$^{-1}$ -- lower
edge).  }
\label{figure:overdensity}
\end{center}
\end{figure}
That this may be the case is hinted at in Fig.\
\ref{figure:overdensity} where we have plotted the surface density of
sources within 50\arcsecs~wide annuli centered on the radio galaxy as
a function of radius from the galaxy. Although the uncertainties are
substantial due to the small number of sources available, we see a
tentative trend showing the SMG surface density increasing towards the
center. The surface densities in the first three inner bins using all
14 SMGs are $\sim$5, 3, and 3$\times$ that of the 1200-$\mu$m blank
field source density (brighter than 2\,mJy -- see Fig.\
\ref{figure:overdensity}), respectively, while we find over-densities
of factors of $\sim 2$ using the MAMBO sources only.

The existence of a SMG over-density around 4C\,41.17 can also be argued
from the fact that if we assume a field surface density of
$N(S_{1200\mu m}\ge 2\,\mbox{mJy}) \sim 700\,$deg$^{-2}$, then we
would expect to find the first mm-source within a radius of
76\arcsecs, the second within 108\arcsecs, the third within
133\arcsecs, etc. What we observe, however, is three sources within
the inner 76\arcsecs, which is thus clearly where the over-density
occurs, while no evidence for an over-density is seen at larger radii.

If we restrict the analysis to the inner 1.5\arcmin, which roughly
corresponds to the radius of the SCUBA map, and consider all five
SCUBA sources within it real, we infer an over-density only 30 per-cent
above that of the blank field, where we have assumed $N(S_{850\mu
m}\ge 2.5\,\mbox{mJy}) \sim 2000\,$deg$^{-2}$ for the latter (Coppin
et al.\ 2006).  Three MAMBO sources and two SHARC-II sources are found
within this radius, corresponding to over-densities of $\sim 4\times$
and $\sim 10\times$, respectively, where we have used $N(S_{350\mu
m}\ge 40\,\mbox{mJy}) \sim 100\,$deg$^{-2}$ from a 350-$\mu$m number
counts model (A.W.\ Blain, private communications) and $N(S_{1200\mu
m}\ge 2.7\,\mbox{mJy}) \sim 380\,$deg$^{-2}$ (Greve et al.\ 2004b).
If we further restrict ourselves to only include J065049.3 and
J065051.4, which are the two brightest SCUBA/MAMBO sources within
1.5\arcmin~of the radio galaxy, we find over-densities of $\sim 7$ and
$\sim 10\times$ above the 850- and 1200-$\mu$m blank field values.
Thus, it is the combination of bright sources close to the radio
galaxy, which makes the over-density significant.

If the SMG over-density depicted in Fig.\ \ref{figure:overdensity} is
correct, it seems to have a scale-length of $\sim 1\arcmin$ (or $\sim
430\,$kpc).  In contrast, the SMG over-density associated with
TNJ1338$-$1942 was more or less evenly distributed over a much larger
area (De Breuck et al.\ 2004). If we compare with observed
over-densities of Ly$\alpha$ emitters towards HzRGs we find typical
scale lengths of 5-10\arcmins, i.e.\ again much larger than the SMG
over-density toward 4C\,41.17.  The one HzRG field (PKS1138$-$262
at $z=2.156$) with H$\alpha$ emitters and EROs (Kurk et al.\ 2004)
shows that these are concentrated within the central 40\arcsecs. It
was suggested that the Ly$\alpha$ emitters are younger, and not
relaxed with respect to the central cluster potential yet, while
EROs and H$\alpha$ emitters were older. Extrapolating this to (sub)mm
galaxies and especially to $z=3.8$ is not straightforward, however.

\bigskip

The above line of arguments is often taken as evidence that 4C\,41.17
as well as other HzRGs (De Breuck et al.\ 2004) are over-dense region
habouring multiple far-IR/(sub)mm luminous systems associated with the
radio galaxy.  Deep SCUBA maps of the central region of 7 HzRGs by
Stevens et al.\ (2003), including 4C\,41.17, yield central
over-densities of SMGs brighter than 5-6\,mJy at 850-$\mu$m of more
than a factor of two associated with these systems. In three cases
they found a companion source brighter than 8\,mJy, corresponding to
over-densities of $\sim 7\times$ that of the field.  A wide 1200-$\mu$m
MAMBO survey of the $z=4.1$ radio galaxy TN\,J1338$-$1942 resulted in
the detection of 10 mm sources within a 25.6 sq.\ arcmin field (De
Breuck et al.\ 2004), corresponding to an over-density $\sim 3\times$
that of the blank field.

In the case of 4C\,41.17, however, which is one of the richest and
complex (sub)mm fields known, we have spectroscopically demonstrated
that at least three of the SMGs in the field, namely J065043.4,
J065049.3 and J065054.0 -- of which the first two were previously
thought to be part of an SMG over-density associated with 4C\,41.17 --
are in fact at much lower redshifts, and therefore completely
unrelated to the radio galaxy.

If we exclude J065043.4, J065049.3 and J065054.0, which we have
demonstrated to lie at redshifts $< 3$, from the analysis, we find
over-densities of $\sim$5, 2, and 2$\times$ that of the 1200-$\mu$m
blank field in the inner three bins in Fig.\ \ref{figure:overdensity}. 
If we further exclude the three sources
(J065040.9, J065055.2, and J065059.6) where the photometric redshifts
indicate $z<3$, we estimate over-densities of $\sim$5, 2, and 1$\times$
that of the 1200-$\mu$m blank field. Thus we find that the evidence
for an over-density of SMGs within the inner 50 and 100\arcsecs~remains
relatively robust with the current, sparse spectroscopic data (but
might disappear with more complete spectroscopy).

It should be noted that while we have been unable to obtain a
spectroscopic redshift for J065051.4, we feel that the likelihood of
this source not being part of the same system as 4C\,41.17 is small,
given its proximity to the radio galaxy and the fact that weak,
extended (sub)mm emission is seen to bridge the two sources (Appendix
A, Fig.\ \ref{figure:4c41-mambo-ps}).  Furthermore, the (sub)mm
photometric redshift of J065051.4 is consistent with this scenario.
Nonetheless, the fact that J065051.4 is the only source for
which there is relatively firm evidence of it belonging to 4C\,41.17,
and the fact this has been shown (via spectroscopy) not to be the case
for three other sources close to 4C\,41.17, illustrates that caution
is needed in the absence of solid spectroscopic evidence when claiming
physical associations between HzRGs and observed SMG over-densities,
especially when the latter are based on small-number statistics.

Based on our spectroscopic and photometric redshifts we conclude that
in the 4C\,41.17 field, nearly half of the apparently associated
SMGs are in fact foreground sources, i.e.\ at $z< 3.8$. This is
fully consistent with the known source density and redshift
distribution of blank field SMGs (Scott et al.\ 2002; Greve et al.\
2004b; Chapman et al.\ 2005). Finally, we note that we find no
spectroscopic evidence for a foreground cluster which may be
gravitationally amplifying the 4C\,41.17 field.

\subsection{The nature of J065043.4}\label{subsection:mambo-blob}

With a flux density of 7.5\,mJy, J065043.4 is by far the brightest
1200-$\mu$m source in the 4C\,41.17 field. What powers the extreme
far-IR luminosity of this source? In a large 1200-$\mu$m survey of the
low-density cluster, Abell\,2125, Voss et al.\ (2006) uncovered four
unusually bright background sources with flux densities similar to
those of J065043.4 ($S_{\rm 1200\mu m} \simeq 10-90\,$mJy).  Three of
these sources were found to be flat-spectrum radio sources and deemed
likely to be quasars based on their X-ray luminosities. These
mm-bright flat-spectrum radio sources were over-dense by a factor 7--9
compared to the expected surface density, derived from the number
counts of flat-spectrum radio sources and a reasonable model for their
behaviour at 90--250\,GHz.  While J065043.4 seems quite similar to the
Voss et al.\ quasars in terms of mm and radio brightness, we are
unable determine whether it is a flat-spectrum radio source because we
lack a meaningful upper limit at 1.4\,GHz (see
\S\ref{subsection:counterparts}). Arguing against J065043.4 being a
flat-spectrum source, such galaxies tend to be type-1 AGN -- bright
optically, and in X-rays -- quite unlike J065043.4.  Further
observations are needed to determine the radio properties of
J065043.4.

The optical spectrum of J065043.4 shows only one very strong
emission line which we identify as Ly$\alpha$ at $z$ = 2.672.  In a 2D
rendition of the spectrum (Appendix A, Fig.\
\ref{figure:spectra-SMGs}a) the Ly$\alpha$ emission is spatially
extended out to 13.5\arcsecs ($\sim 110\,$kpc) and there is evidence
of a velocity shear across the line, which could be due to either
rotation or infall/outflow.

Extended Ly$\alpha$ haloes have been observed around a variety of
objects at high redshifts, including HzRGs (e.g.\ Reuland et al.\
2003) and SMGs (Ivison et al.\ 1998; Chapman et al.\ 2004). Deep
narrow-band imaging surveys specifically designed to select such
`Ly$\alpha$ Blobs' (LABs) have been undertaken (e.g.\ Fynbo, M{\o}ller
\& Warren 1999; Steidel et al.\ 2000). Nevertheless, LABs such as
J065043.4, with Ly$\alpha$ emission extending beyond $50\,$kpc, are
very rare; only a dozen or so have been found.  It is unclear what
powers these very extended LABs. Plausible explanations are a
large-scale superwind stemming from a massive, dust-enshrouded
starburst (Taniguchi \& Shioya 2000; Ohyama et al.\ 2003), a buried
QSO (Haiman \& Rees 2001; Weidinger et al.\ 2005), or simply cooling
radiation from infalling gas within a dark matter halo (e.g.\ Haiman
et al.\ 2000).

J065043.4 is undetected at 24-$\mu$m -- unusual but not unique for an
SMG; its single emission line, Ly$\alpha$, is perfectly normal
(Chapman et al.\ 2005). It is the third known submm-bright LAB and all
three have different properties -- typical of this intriguing,
heterogeneous population (e.g.\ Ivison et al.\ 2000). A more detailed
study of J065043.4 is in progress (Greve et al.\ in prep.).

\section{Summary}\label{section:sumary}

We have conducted wide-field imaging at (sub)mm, mid-IR and radio
wavelengths of the luminous, $z=3.792$ radio galaxy, 4C\,41.17. In
addition to the radio galaxy itself, we robustly confirm two bright
sources previously detected by SCUBA in its immediate vicinity. Of
another three very faint sources detected by SCUBA, we tentatively
confirm one at 350-$\mu$m. A further nine sources are detected (at the
3.5-$\sigma$ level) at 1200-$\mu$m in this field.  Thus a total of 14
(sub)mm galaxies are found within a $\sim 4.2\arcmin$ radius of
4C\,41.17.

Using our radio and 24-$\mu$m data we find statistically significant
counterparts to 8/14 (57 per-cent) of our SMGs. We targeted six of
these robust counterparts spectroscopically using LRIS on Keck, and
were able to infer reliable spectroscopic redshifts in four cases. All
four were found to lie at redshifts well below that of the radio
galaxy, thus ruling out any physical association with 4C\,41.17.
Comparing the spectroscopic redshifts with redshifts derived from the
1.6-$\mu$m stellar feature suggest that the latter is a reliable
photometric redshift indicator. Of four SMGs with no spectroscopic
redshifts we use the 1.6-$\mu$m stellar bump to estimate their
redshifts and find only one to be consistent (within the errors) with
the redshift of 4C\,41.17.

Armed with our spectroscopic redshifts, we find that nearly half of the
SMGs apparently associated with 4C\,41.17 are in fact foreground
sources. We have constrained the over-density of SMGs within a
50\arcsecs~region of 4C\,41.17 to be $\sim 5\times$ that of the
typical surface density observed in blank fields. This drops off
as a function of distance from the radio galaxy, with an apparent
scale-length of $\sim 1\arcmin$.

Deeper, wider (sub)mm surveys of of HzRGs are needed to improve the
statistics on source over-densities and to probe the filamentary
structures expected to be channeling material into these
regions. These surveys should be complemented by high-resolution
(sub)mm interferometry and deep radio imaging to pin-point positions
determine source morphologies. Finally, deep spectroscopy is required
to test for membership of a HzRG-related proto-cluster, although the
imminent arrival of broad-band heterodyne receivers suggests CO
spectroscopy may prove to be a more efficient way of determining
redshifts for these SMGs.

\section*{Acknowledgments} We are grateful to Darren Dowell and 
Colin Borys for helpful discussion relating to SHARC-II observations
and data reduction.  We thank Andrew Blain for providing us with
350-$\mu$m number counts predictions. We are also grateful to Jason
Stevens and Nick Seymour for helpful suggestions. We thank the IRAC
instrument team for allowing publication of GTO data on 4C\,41.17. We
also thank Mark Dickinson and Emily MacDonald for providing the
DEIMOS/Keck $z$-band image of 4C\,41.17.  Some of the data presented
herein were obtained at the W.M.\ Keck Observatory, which is operated
as a scientific partnership among the California Institute of
Technology, the University of California and the National Aeronautics
and Space Administration. The Observatory was made possible by the
generous financial support of the W.M.\ Keck Foundation. The authors
wish to recognize and acknowledge the very significant cultural role
and reverence that the summit of Mauna Kea has always had within the
indigenous Hawaiian community; we are most fortunate to have the
opportunity to conduct observations from this mountain. The work of DS
was carried out at Jet Propulsion Laboratory, California Institute of
Technology, under a contract with NASA.

\appendix{\noindent\bf{APPENDIX A: NOTES ON INDIVIDUAL SOURCES}}\\
\noindent

%
%
\begin{figure*}
\begin{center}
\includegraphics[width=0.9\hsize,angle=0]{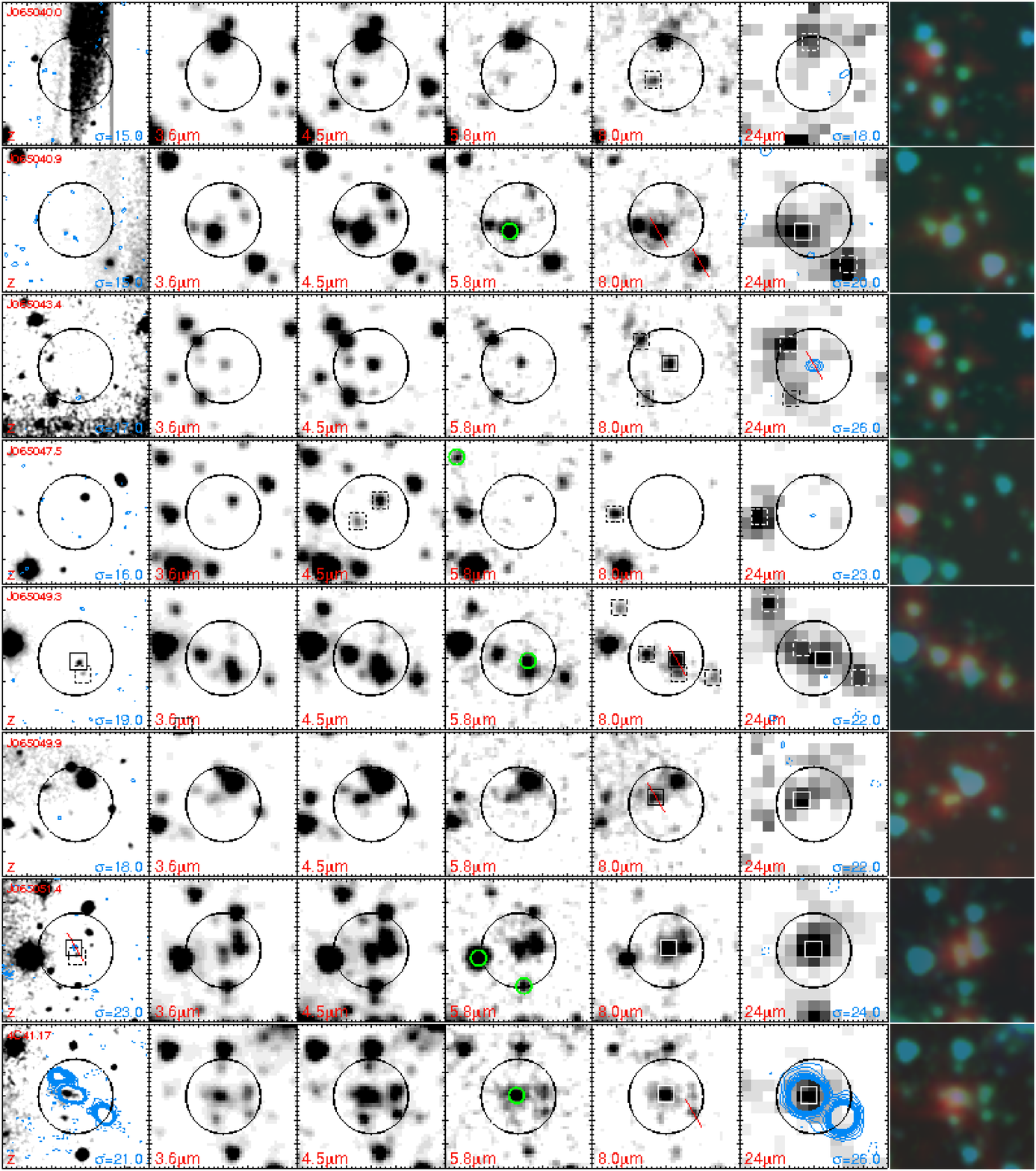}
\caption[]{$30\arcsecs \times 30\arcsecs$ postage stamp images of the
14 SMGs (and 4C\,41.17) detected towards 4C\,41.17. The postamp images
are centered on the (sub)mm centroid. Although, in cases where a
source was detected at more than one (sub)mm wavelength, the postamp
image has been centered on the average (sub)mm centroid (Table
\ref{table:data}).  The leftmost column shows $z$- band images
(grey-scale) with radio 1.4-GHz contours overlaid.  The radio contours
are plotted at $-3,3,4~...~10, 20~...~100\times \sigma$, where
$\sigma$ is the local rms noise. The middle five columns (from left to
right) show grey scale images at 3.6-, 4.5-, 5.8-, 8.0-$\mu$m (IRAC)
and 24$\mu$m (MIPS). The latter with 4.9-GHz radio contours overlaid
at $-3,3,4~...~10, 20~...~100\times \sigma$, where $\sigma$ is the
local rms noise. The solid central circle indicates the $8\arcsecs$
error circle within which we have searched for radio and mid-IR
counterparts. Solid squares indicates robust identifications (i.e.\
$P\le 0.05$) based on radio or 24-$\mu$m counts, while dashed squares
indicate possible counterparts. Red lines represent the position of
slits on our LRIS masks and green circles represent X-ray sources
detected with Chandra (Smail et al.\ 2003b). The last column shows the
'true' color postamp images, based on the 3.6-$\mu$m (blue),
4.5-$\mu$m (green) and 24-$\mu$m (red) images.}
\label{figure:postamps}
\end{center}
\end{figure*}
%
%
\begin{figure*}
\setcounter{figure}{0}
\begin{center}
\includegraphics[width=0.9\hsize,angle=0]{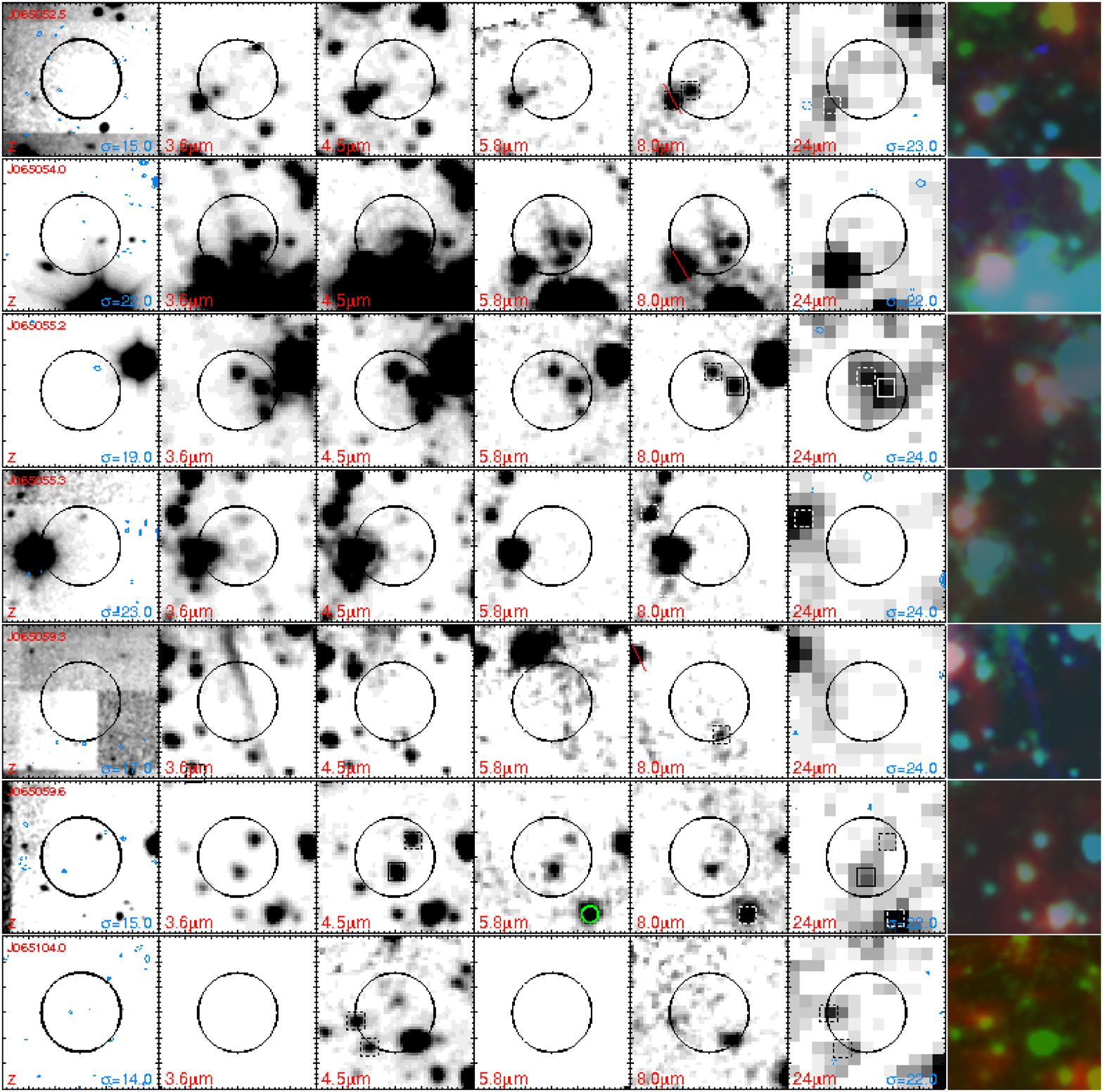}
\caption[postamps]{continued... 
}
\label{figure:postamps-1}
\end{center}
\end{figure*}

\begin{figure*}
\begin{center}
\includegraphics[width=0.9\hsize,angle=0]{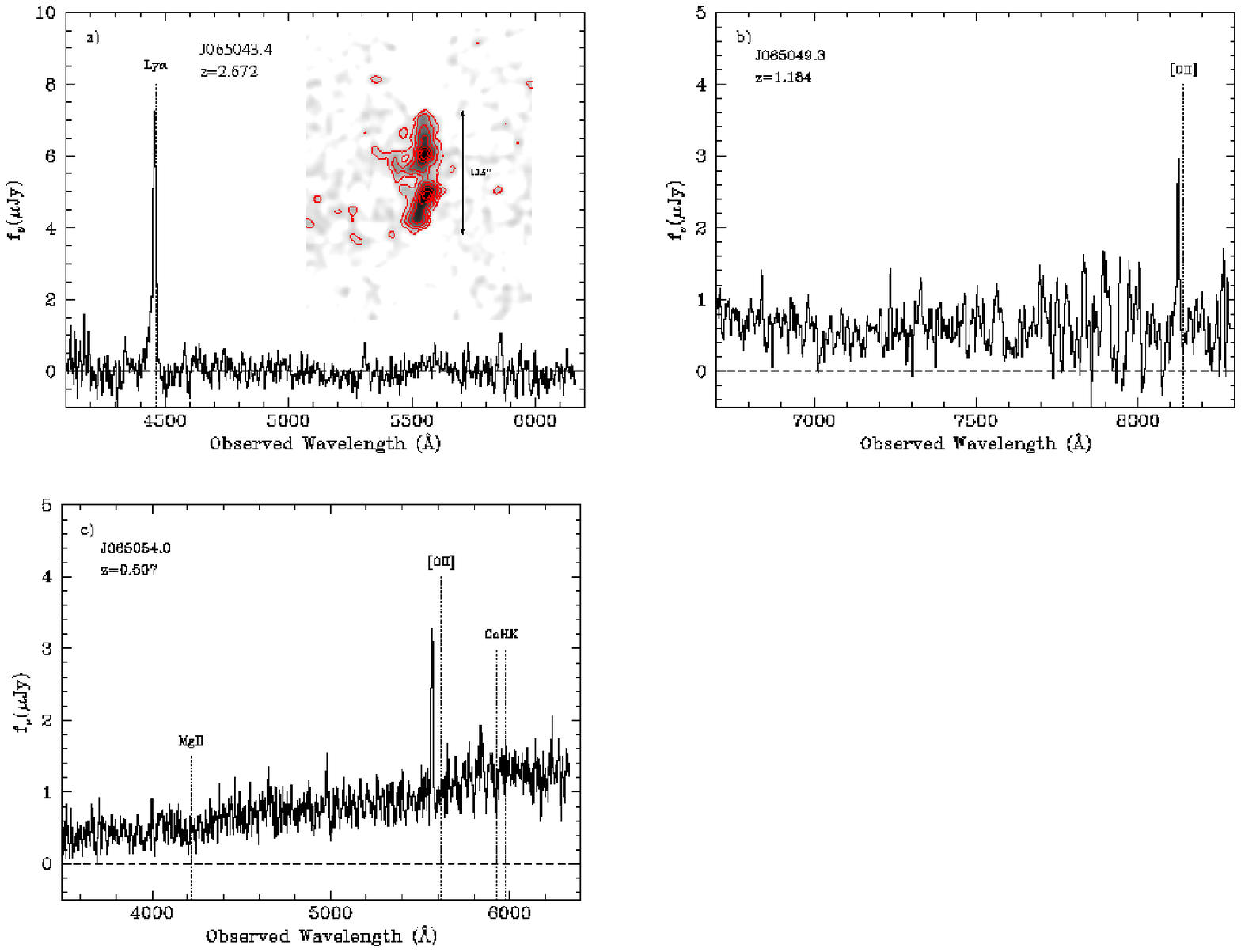}
\caption[postamps]{Keck/LRIS spectra of J065043.4 (a), J065049.3 (b)
and J065054.0 (c). The insert gray-scale 2D spectrum in a) shows the
extended Ly$\alpha$ emission associated with J065054.0. The spatial
axis is vertical and the spectral axis is horizontal.  }
\label{figure:spectra-SMGs}
\end{center}
\end{figure*}

\begin{figure}
\begin{center}
\includegraphics[width=0.9\hsize,angle=-90]{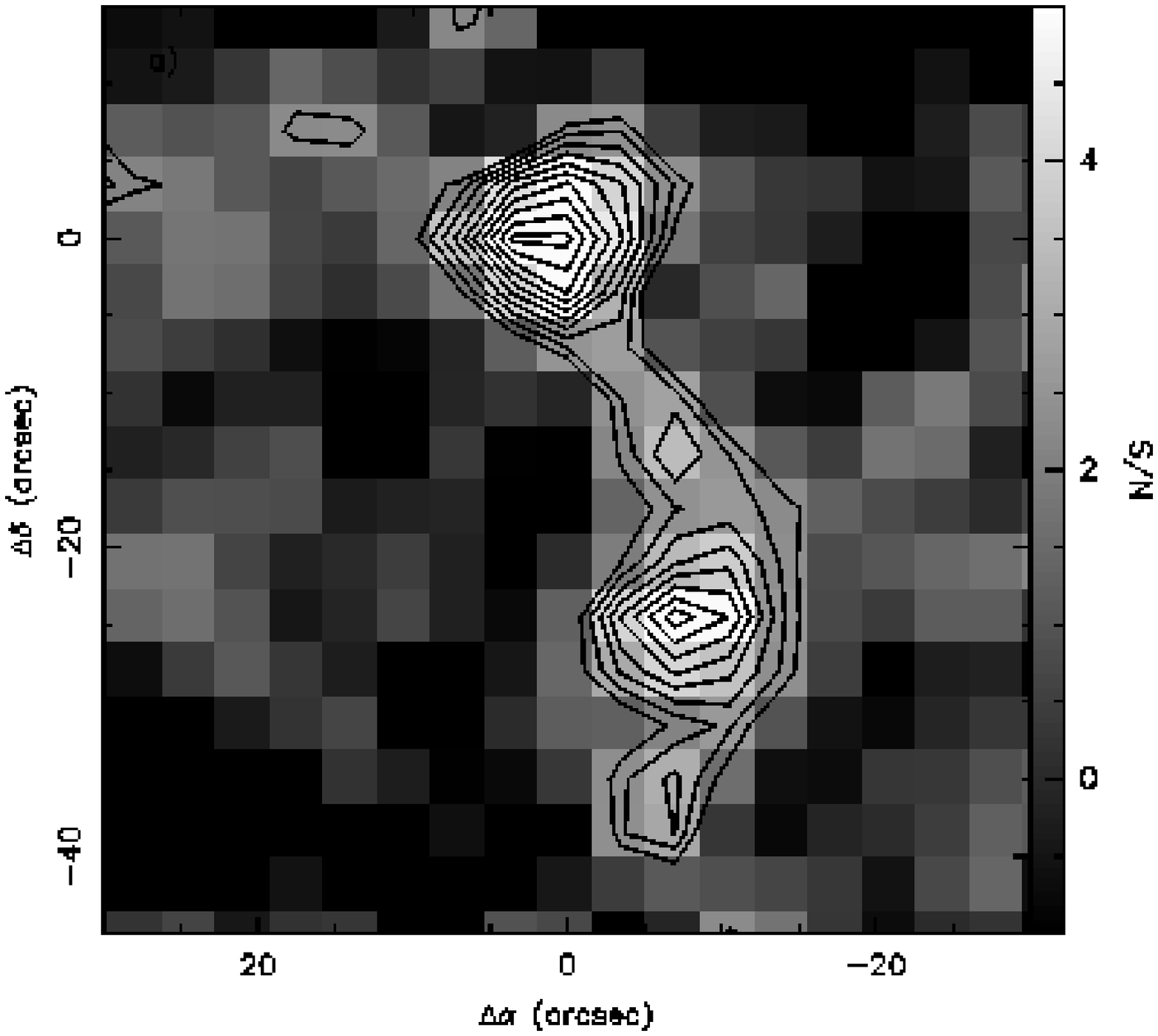}
\caption[4c41-mambo-ps]{A zoom-in of the inner $30\arcsecs \times
30\arcsecs$ region around 4C\,41.17 and J065051.4 in the 1200-$\mu$m
signal-to-noise map. Contours are at SNR=2.0,2.5,3.0,...,8.0.  }
\label{figure:4c41-mambo-ps}
\end{center}
\end{figure}

\noindent
{\bf J065040.0:} A radio-blank field containing two possible mid-IR
counterparts. One is detected in the IRAC bands only and is located
3\arcsecs~south-east of the SMG position.  The other, located about
7\arcsecs~north of the (sub)mm centroid is also strongly detected in
the near-IR and at 24-$\mu$m. The likelihood of either source being a
chance associations is non-negligible, however, and we are therefore
unable to unambiguously identify the correct near-/mid-IR counterpart
for this SMG.\\
\noindent
{\bf J065040.9:} A radio-blank field, but several possible
counterparts are seen in the IRAC bands.  The two brightest 8.0-$\mu$m
sources coincide with an extended, apparently blended 24-$\mu$m source
about 5\arcsecs~south-east of the (sub)mm centroid. The 24-$\mu$m
peak, however, coincides with the brightest IRAC source suggesting
that most of the 24-$\mu$m emission comes from this source. Invoking
the findings by Pope et al.\ (2006) that in most cases where multiple
24-$\mu$m sources are found within a SCUBA beam, only one -- namely
the brightest 24-$\mu$m source -- is dominating the submm emission, it
would suggest that this source is the counterpart to the SMG.  This is
further strengthened by the fact that the source also has a X-ray
counterpart detected by Chandra. We find the 24-/1200-$\mu$m
association to be statistically significant ($P<0.029$). The IRAC
source was targeted spectroscopically but due to its faintness in the
near-IR and the poor quality of the spectrum we were unable to infer a
redshift.  The spectrum of a second 24-$\mu$m source lying
11.5\arcsecs~to the south-west of the mm centroid yielded a redshifts
of $z=0.909\pm0.002$. This source, however, is not a statistically
significant counterpart.\\
\noindent
{\bf J065043.4:} Two 24-$\mu$m sources are found at the edge of the
error circle, but neither of them is a statistically robust
counterpart (see Table 4). However, a statistically significant
mm/radio association ($P=9\times 10^{-5}$) is obtained for the robust
(4.5-$\sigma$) 4.9-GHz source ($S_{4.9\rm GHz}=109.2\pm 31.1\,\mu$Jy)
near-coincident with the 1200-$\mu$m centroid. The radio source,
which appears to be unresolved, coincides with a point-source detected
in all the IRAC bands, but no counterpart is seen in the $z$ band nor
at 24-$\mu$m. Based on the strong 4.9-GHz radio ID, we adopted this
source as the radio/mid-IR counterpart to J065043.4 and it
was subsequently targeted with LRIS. The resulting spectrum is shown
in Fig.\ \ref{figure:spectra-SMGs}a: strong Ly\,$\alpha$ emission,
extended on scales of 13.2\arcsecs~is detected at a
wavelength corresponding to $z=2.672\pm0.001$.\\
\noindent
{\bf J065047.5:} A strong 24-$\mu$m source is located $\sim
11\arcsecs$ east of the mm centroid, but does not represent a
statistically significant identification. Two IRAC sources are found
within the positional error circle of this radio-blank field, but
neither source is associated with any 24-$\mu$m emission and their
association with the SMG is not significant.  Thus, the correct
mid-IR/radio counterpart for this source is undetermined.\\
\noindent
{\bf J065049.3:} Two faint $z$-band sources, separated by $\sim
2.7\arcsecs$, and their matching counterparts in the IRAC channels are
detected near the center of the error beam. The two source (K1 and K2)
were initially suggested as near-IR counterparts to the SMG by Ivison
et al.\ (2000) on the basis of their resolved 850-$\mu$m emission (see
Fig.\ 1 of Ivison et al.\ 2002b) -- with K1 denoting the source
closest to the center. K1 is the reddest, while K2 becomes fainter and
more extended in the long-wavelength IRAC channels. In the 24-$\mu$m
MIPS band the two sources, together with an IRAC source $\sim
5\arcsecs$ west of the center, combine into a single, blended source.
We note that K1, apart from being the reddest IRAC source and the only
X-ray source within the error circle, also coincides with the peak of
the (blended) 24-$\mu$m emission. This peak is dominating the total
flux, thus making it the most likely mid-IR counterpart to the submm
emission. In fact, we find a statistically significant 1200-/24-$\mu$m
association ($P=0.011$) for K1. Moreover, LRIS spectroscopy of K1
yields a redshift of $z=1.184\pm0.002$ (Fig.\
\ref{figure:spectra-SMGs}b), which is within the redshift distribution
of blank-field SMGs (Chapman et al.\ 2005) thus further suggesting
that this is the correct counterpart. It is possible, however, that K1
is part of a composite system which include the bluer component
K2. From the estimated 24-$\mu$m flux of the third IRAC source within
the error beam (Table 4), we find this to be a non-significant
24-$\mu$/mm association ($P=0.067$). A further two 24-$\mu$m sources
are found $15.0\arcsecs$ south-west and $10.5\arcsecs$ north-east of
the mm centroid - but neither qualify as robust counterparts.  Thus we
conclude that K1 is the correct near-/mid-IR counterpart to J065049.3
with K2 being a possible associated system. \\
\noindent
{\bf J065049.9:} while no radio sources are detected within
8\arcsecs~of the 1200-$\mu$m centroid, at least five IRAC sources are
found within the search radius. The reddest ($\log (S_{5.8\mu
m}/S_{3.6\mu m} = -0.045$) IRAC source, which is also the one closest
to the mm centroid, coincides with a robust 24-$\mu$m source. The
latter constitutes a robust 1200-/24-$\mu$m association ($P=0.012$)
and we take this source to be the mid-IR counterpart to
J065049.9. LRIS spectroscopy of this source did not yield any
discernable lines/continuum.\\
\noindent
{\bf J065051.4:} an unambiguous mm/radio association ($P=0.002$) is
found from the 1.4-GHz map which shows a robust source at R.A.:
06:50:51.4 and Dec.: +41:30:06.5, i.e.\ within $\sim 1\arcsecs$ of the
1200-$\mu$m centroid. The radio source coincides with the reddest
($\log (S_{5.8\mu m}/S_{3.6\mu m} = 0.23$) and most northern component
of a slightly blended IRAC source as well as with a robust 24-$\mu$m
point-source ($S_{24\mu m}=448\pm 24\,\mu$Jy). The latter yields a
robust 1200-/24-$\mu$m association ($P=0.002$).  A faint $z$-band
source is seen just $\sim 1\arcsecs$ south of the radio source. This
source is also detected at 3.5- and 4.5-$\mu$m where it appears to be
interacting with the radio-identified IRAC source. Given their
proximity and merger-like appearance it is tempting to speculate that
both sources are part of the same system. Unfortunately, the optical
LRIS spectrum, obtained with a slit at the radio position, did not
yield any identifiable line or continuum emission; we are therefore
unable to infer a spectroscopic redshift for this source.  At a
resolution of $3.1\arcsecs \times 2.3\arcsecs$, continuum observations
at 1.3\,mm of this source with the IRAM Plateau de Bure Interferometer
(PdBI) failed to yield a detection (3-$\sigma \ls 2.2\,$mJy) (D.\
Downes, private communications). The blank IRAM PdBI image implies
that the (sub)mm emission is spread over a large area or breaks down
into two or more fainter components. Based on present data the
evidence is ambiguous, although there may be some support for the
latter scenario looking at Fig.\ \ref{figure:4c41-mambo-ps} where we
have zoomed in on a $30\arcsecs \times 30\arcsecs$ region around
4C\,41.17 and J065051.4. Contours at the $2-3$-$\sigma$ level show
evidence of faint emission extending between the two sources as well
as the presence of two faint sources north and south of J065051.4.  \\
\noindent
{\bf 4C\,41.17} The two radio-lobes and the radio galaxy itself are
evident from the 1.4-GHz radio contours.  The radio galaxy coincides
with the (sub)mm centroid and is clearly detected in the $z$-, IRAC-,
and 24-$\mu$m bands as well as in the X-rays. The south-western radio
lobe coincides with a faint source detected in the IRAC bands. This
source was first detected at 2.15-$\mu$m ($K_s$-band) by Papadopoulos
et al.\ (2005) who speculated it could be due to jet-induced star
formation.  The source was targeted spectroscopically but
unfortunately no discernable line nor continuum emission was
detected.\\
\noindent
{\bf J065052.5} A radio-blank field, in which a robust 24-$\mu$m
source is seen $\sim 9\arcsecs$ to the south-east of the mm centroid,
just outside the positional error circle. It does not, however,
represent a significant 24-$\mu$m/mm association ($P=0.084$). The LRIS
spectrum of this source spectrum did not yield any discernable
emission lines. An IRAC source about 5\arcsecs~south-east of the mm
centroid, which coincides with extremely faint $z$-band emission but
is undetected at 24-$\mu$m, is another possible, but not statistically
significant, counterpart. The radio/mid-IR identification of this SMG
therefore remains ambiguous.  \\
\noindent
{\bf J065054.0} No robust radio sources are detected in this
field. Three IRAC sources lie within the 8\arcsecs~search circle,
neither of which qualify as robust counterparts to the SMG.  However,
a strong 24-$\mu$m source $\sim 9\arcsecs$ south-west of the mm
centroid is a statistically significant mid-IR 24-/mm association ($P
= 0.043$). Even though this source lies (just) outside of the
8\arcsecs~search radius we still adopt it as the correct 24-$\mu$m
counterpart. In the $z$-band the source appears as a lenticular
galaxy. The Keck/LRIS spectrum of this source is shown in Fig.\
\ref{figure:spectra-SMGs}c and we infer a spectroscopic redshift of
$z=0.507\pm 0.002$.  \\
\noindent
{\bf J065055.2} Bright, extended 24-$\mu$m emission is found within
the 8\arcsecs~search radius in this radio-blank field. In the IRAC
bands this emission is resolved into three separate sources.  The bulk
of the 24-$\mu$m emission coincides with the brightest IRAC source
located $\sim 2.9\arcsecs$ norh-west of the mm centroid, and we find
the probability of this being a non-random association to be
statistically significant ($P=0.031$). We have estimated the
contribution to the 24-$\mu$m emission by the two other IRAC sources
and find that they both fall short of being a robust 24-$\mu$m/mm
association.\\
\noindent
{\bf J065055.3} A radio-blank field which is also hard to interpret in
the mid-IR.  A bright 24-$\mu$m source lies $\sim 14\arcsecs$
north-east of the mm centroid but does not make the cut as a robust
counterpart ($P=0.098$). Two very faint $z$-band source are detected
$\sim 6\arcsecs$ north and $\sim 3\arcsecs$ south-west of the SMG
centroid. Both sources have weak counterparts in the 3.6- and
4.5-$\mu$m IRAC channels, but elude detection at 5.8- and 8.0-$\mu$m
as well as at 24-$\mu$m. Neither source represent a significant
mid-IR/mm association, and the correct counterpart for this SMG
therefore remains elusive.  \\
%
%
\noindent
{\bf J065059.3} A radio-blank field. A faint, morphologically complex
IRAC source $\sim 7\arcsecs$ south-west of the SMG position is the
only source found within the 8\arcsecs~search radius. It not detected
at 24-$\mu$m nor in the near-IR, and it does not constitute a
statistically significant mid-IR/mm association.  Thus we are a not
able to assign a robust mid-IR identification to this SMG.\\
%
%
\noindent
{\bf J065059.6} An robust 24-$\mu$m source peaks $\sim$4\arcsecs~south
of the mm centroid and makes up a statistically significant
24-$\mu$m/mm association ($P=0.048$). The source coincides with an
IRAC source with a slightly extended appearance. Extremely faint
24-$\mu$m emission is located $\sim 6\arcsecs$ to the north-west of
the center via its IRAC and $z$-band counterparts, while a bright,
X-ray detected 24-$\mu$m source is seen $\sim 14.8\arcsecs$ from the
center. However, neither of these two sources constitute an
unambiguous identification. \\
\noindent
{\bf J065104.0} Guided by the IRAC images, two faint 24-$\mu$m source
are found $\sim 9\arcsecs$ east and south-east of the mm
centroid. However, neither source is a robust counterparts ($P=0.078$
and $P>0.1$), respectively, and the correct identification therefore
remains ambiguous.  \\

\bsp

\label{lastpage}

\end{document}